\documentclass[reprint,aps,prl,showpacs]{revtex4-2}

\usepackage[utf8]{inputenc}
\usepackage{graphicx}
\usepackage{tikz}
\usepackage{dcolumn}
\usepackage{amsmath}
\usepackage{amsfonts}
\usepackage{amssymb}
\usepackage{hyperref}
\usepackage{pbox}
\usepackage{pinlabel}
\usepackage{ragged2e}
\usepackage[justification=RaggedRight]{caption}
\usepackage[singlelinecheck=off]{subcaption}
\usepackage{ulem}

\begin{document}

\title{Minimal theory of strange carriers}

\author{Simone Fratini}
\affiliation{Universit\'e Grenoble Alpes, CNRS, Grenoble INP, Institut N\'eel, 38000 Grenoble, France}

\begin{abstract}
I explore a theory of transport and optical properties of strange metallic carriers in strongly correlated systems that follows from assuming that the diffusion constant has reached its quantum limit $D=\hbar/m$, 
and that such quantum carriers behave as distinguishable particles as they would in an electronic solid. These assumptions immediately lead to $T$-linear resistivities with apparent Planckian scattering rates and, extending to the frequency domain, to the stretched Drude peaks and $\omega/T$ scaling commonly observed in optical absorption experiments in strange metals.
This behavior can be rationalized by observing that when the thermal de Broglie length $\lambda_{dB}$ exceeds the mean-free-path,  the carrier motion can no longer be described in terms of random collisions of classical particles as assumed by Drude-Boltzmann theory and should be viewed instead as a sequence of projective measurements collapsing the wavefunction. 
\end{abstract}

\maketitle

\paragraph{Introduction.---}

Strange metals
are  characterized by $T$-linear resistivities extending over broad doping and temperature ranges. 
When interpreted in terms of the classical Drude theory of metals, $\rho=1/\sigma_D=1/(ne^2\tau/m)$, these yield universal Planckian scattering rates $\hbar/\tau \simeq k_B T$ \cite{PhillipsScience22,Bruin,Legros}. 
Ubiquitous in strange metals is also the observation that transport anomalies extend to the frequency domain: the optical conductivity $\sigma(\omega)$
exhibits a power-law decay that is slower than the $1/\omega^2$ expected from simple exponential relaxation of the momentum
\cite{PhillipsScience22,vdMarelNature03,MichonNatComm23,vanHeumen22} and approximate $\omega/T$ scaling. 
As the whole function of frequency 
fundamentally deviates from the Drude form, one may wonder if we are still allowed to use the same Drude theory to describe the temperature dependence of its d.c. limit $\rho=1/\sigma(\omega\to 0)$, or if a thoroughly different description is needed.  In that case, the question is what is the meaning of an apparently Planckian scattering rate.

\begin{figure}
    \centering
    \includegraphics[width=8.5cm]{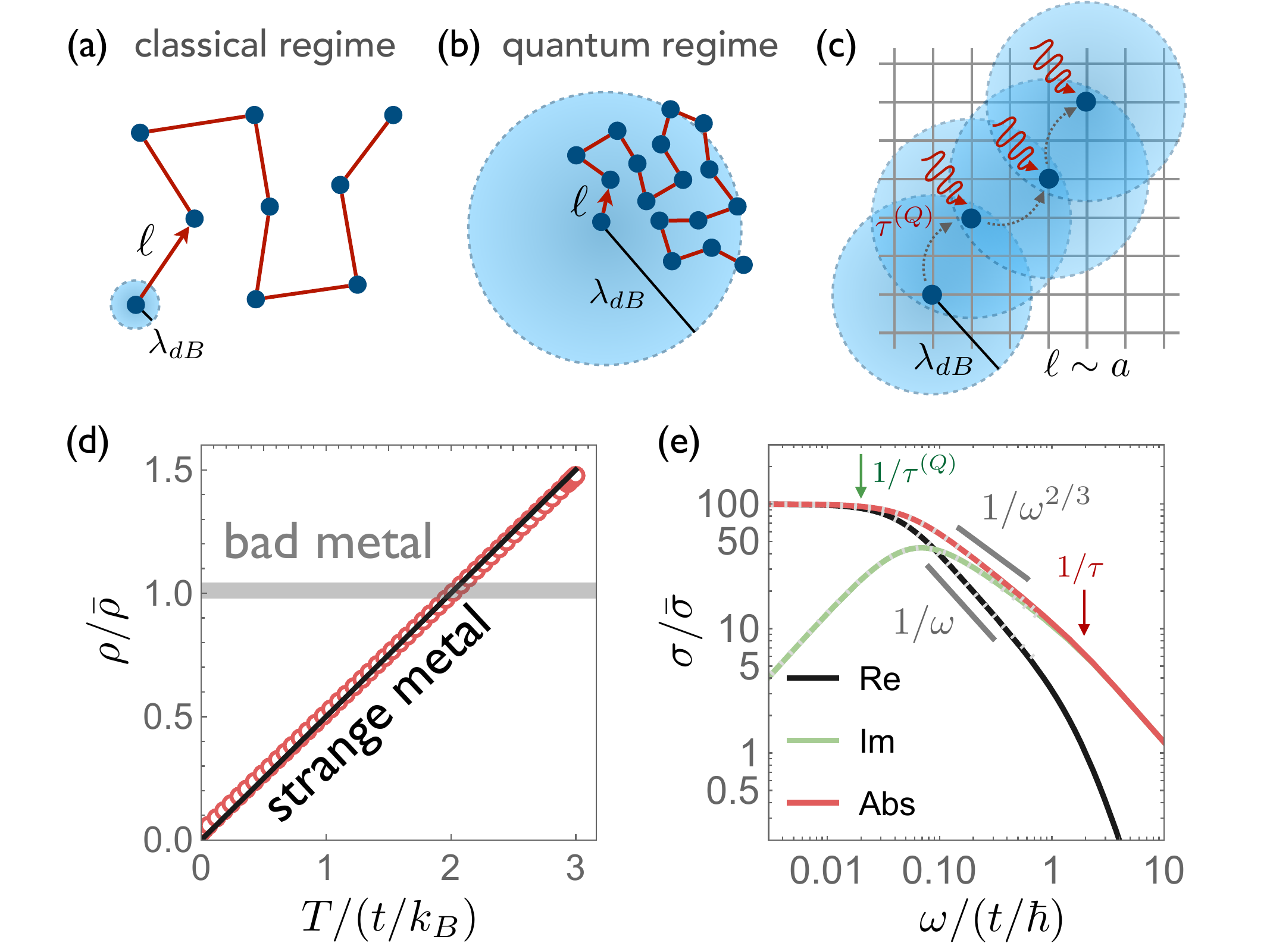}
    \caption{\textbf{Transport of strange carriers.} 
    (a) Diffusion of classical particles traveling ballistically with velocity $v$ and undergoing random collisions with a mean-free-path $\ell$, corresponding to a Drude-like behavior of the conductivity. (b) When the de Broglie length $\lambda_{dB}$ is larger than the mean-free-path, the classical picture breaks down due to the increasing quantum uncertainty. (c) Charge transport can be viewed as a series of projective measurements allowing the quantum particle to diffuse over a length $\lambda_{dB}$ over a Planckian timescale $\tau^{(Q)}$, corresponding to a diffusion constant $D=(\lambda_{dB})^2/\tau^{(Q)}=\hbar/m$. (d) The resulting resistivity is linear in temperature, falling below the classical value set by the Mott-Ioffe-Regel limit. The numerical result for a weakly doped Mott insulator in the limit $U/t \to \infty$ \cite{Planckian-tJ} is reported as open circles. (e) The corresponding optical conductivity features a stretched Drude peak whose width is governed by the same Planckian scale $\tau^{(Q)}$ (dashed lines: see note \cite{noteCD}).}
    \label{fig:sketch}
\end{figure}

\paragraph{Drude theory and its limitations.---}
The Drude  theory of transport considers classical particles traveling ballistically with velocity $v$ and undergoing random collision events that cause the momentum to relax over a typical timescale $\tau$ (the scattering time), or a typical lengthscale $\ell=v\tau$ (the mean-free-path), illustrated in Fig. \ref{fig:sketch}(a). 
The resulting conductivity can be obtained from Einstein's relation, $\mu = (e/k_BT) D$, describing the mobility of independent classical particles of mass $m$ with a diffusion constant $D=v^2\tau$ and a thermal velocity set by equipartition, $v^2=k_BT/m$. Using $ \sigma=ne\mu$  and collecting the factors leads to $\sigma_D=ne^2\tau/m$.  

The same equation also applies to metals under certain conditions. While the above derivation clearly does not hold for degenerate electrons, as the semi-classical Boltzmann transport equation combined with Fermi-Dirac statistics implies that the conductivity is solely determined by the electrons near the Fermi energy \cite{Greenwood58} --- with density of states $N(E_F)$ and diffusivity $D_F=v_F^2\tau$ --- not by the total density $n$, the Drude result is accidentally recovered when the 
band dispersion is strictly parabolic \cite{Ashcroft}. In real solids, although not strictly correct, the Drude formula is still viewed as a useful approximation owing to its simplicity.

At any rate, any semi-classical description
can only hold if the quantum nature of the electronic carriers can be neglected on the scale of the microscopic processes involved. The usual statement is that the scattering rate $\hbar/\tau$ should be small compared to the thermal scale $k_BT$ \cite{Greenwood58,Ashcroft}. 
More useful to our purposes, this is equivalent to saying that the thermal de Broglie length $\lambda_{dB}=\hbar/\sqrt{mk_BT}$ measuring quantum uncertainty should be much shorter than the mean-free-path $\ell=\sqrt{k_BT/m}\tau$, ensuring that the charge carriers can be effectively viewed as classical objects (Fig. \ref{fig:sketch}(a)).
What happens then in bad metals, where $\ell$ can be as short as the inter-atomic distance $a$ on a lattice, easily leading to the opposite situation
$\lambda_{dB} \gg \ell$? (Fig. \ref{fig:sketch}(b))

\paragraph{Renouncing the Drude paradigm.---}
We know from the Schr\"odinger equation that the natural unit of diffusion for quantum particles is 
\begin{equation}
    D^{(Q)}=\hbar/m.
    \label{eq:Dq}
\end{equation}
Let us assume that the carriers in bad metals have reached this quantum lower bound \cite{Sommer11,Bardon14,Planckian-tJ,AydinPNAS23}.
\footnote{On a lattice, the lower bound Eq. (\ref{eq:Dq}) is equivalent to the Mott-Ioffe-Regel (MIR) condition  stating that wavelike coherence is lost at each hop: setting $\ell=a$ with $\ell=v \tau$  the  mean-free-path and $a$  the nearest-neighbor distance, and taking $v=2 t a/\hbar$ describing ballistic motion of electrons with  transfer integral $t$
yields a scattering rate $\tau_{MIR}=\hbar/2t$ and a  diffusion constant $D=2t a^2/\hbar$. Using   $m= \hbar^2/(2ta^2)$  leads to $D^{(Q)}=\hbar/m$.}
Applying again the Einstein relation  we obtain  
$\sigma^{(Q)}(T)=ne^2 (\hbar/k_BT)/m $. It therefore immediately appears that if interpreted in the framework of Drude theory, this result  corresponds to an apparent scatterting time
\begin{equation}
    \tau^{(Q)}= \frac{\hbar}{k_BT}.  
    \label{eq:Planckiantime}
\end{equation}
The corresponding resistivity is  
\begin{equation}
    \rho^{(Q)}(T)=\frac{mk_B}{ne^2\hbar} \ T,
    \label{eq:Planckianrho}
\end{equation}
which explains the temperature-dependence observed experimentally. Numerically, Eqs. (\ref{eq:Dq}) and (\ref{eq:Planckianrho}) have been shown to hold quantitatively in the t-J model due to 
the presence of strong spin fluctuations  \cite{Planckian-tJ} (see also Ref. \cite{Kokalj} on the Hubbard model). Alternatively, the presence of abundant slow lattice vibrations has been proposed to lead to the same result \cite{AydinPNAS23}, highlighting the generality of the phenomenon. Arguably, quantum critical fluctuations, or the scattering to other emergent collective modes, could also achieve the same scope \cite{Grilli23}.

It may not have escaped the reader that 
the use of Einstein's relation itself constitutes a second, strong assumption, being an implicit statement of particle distinguishability. 
While this is expected to hold in strongly correlated systems near half-filling, where the effective number of carriers is low and the carriers are non-degenerate, the same is not granted at large dopings and low temperatures. 
One may still argue that in the presence of strong correlations classical statistics could persist in some form,
because the spatial structure of strongly correlated liquids closely resembles that of solids, where particle exchange  becomes irrelevant \cite{Mahan}. We shall 
come back to this point 
in the conclusions.

\paragraph{Collisions as projective measurements.---}
As the concept of collisions of ballistic classical particles loses meaning and the Drude formula does not apply, \textit{how should we then interpret  Eq. (\ref{eq:Dq}) and the resulting Planckian time Eq. (\ref{eq:Planckiantime})?} Because of quantum uncertainty, the carrier wavefunctions effectively extend over a large radius $\lambda_{dB}$. We can be tempted to replace the concept of classical collisions with that of projective measurements, events as a result of which the wavefunction itself can collapse at any point within this radius. Carrier diffusion would then occur through subsequent expansion and collapse of the wavefunction \cite{LiFischerPRB18}, as sketched in Fig. \ref{fig:sketch}(c).  In this framework it is now natural to interpret $\tau^{(Q)}$ as the typical time it takes to thermally collapse the wavefunction: 
the diffusion constant  of particles traveling over a distance $\lambda_{dB}$ at every $\tau^{(Q)}$ is precisely $D=(\lambda_{dB})^2/\tau^{(Q)}=\hbar/m$, i.e. Eq. (\ref{eq:Dq}). The microscopic transport mechanism depicted here bears some resemblance with transient localization \cite{Ciuchi11,Fratini-AdvMat16, Giannini}, that describes the breakdown of Drude theory in organic semiconductors. There, localized states (radius $L$) diffuse following the dynamics of molecular disorder (timescale $\tau_d$), with a diffusion constant $D=L^2/\tau_d$ replacing the semi-classical $v^2\tau$.

We note that if we took  Drude theory literally, the bad metal condition $\ell\sim a$ (or, equivalently,  $\hbar/\tau \sim t$ with  $t$ the hopping integral) would imply a large, constant resistivity of the order of the Mott-Ioffe-Regel (MIR) value. On a layered two-dimensional lattice with band mass $m= \hbar^2/(2ta^2)$, interlayer distance $d$ and concentration $x$ per unit cell, this is equal to the resistivity unit $\bar\rho \equiv (\hbar/e^2) d/x$. Quantum uncertainty allows the carriers to diffuse faster than what is allowed by classical mechanics:  as a result, what would classically be a bad metal with a large resistivity $\bar\rho$ is converted into a strange metal with a much lower resistivity given by Eq. (\ref{eq:Planckianrho}) or, in lattice units,  $\rho^{(Q)}(T)=\bar\rho \ \frac{k_BT}{2t}$ (Fig. \ref{fig:sketch}(d)).

\paragraph{Strange magnetoresistance.---}
One of the open experimental puzzles in strange metals is the observation of  $B/T$ scaling in magneto-transport  \cite{Ayres21}. To understand the effect of a magnetic field on the diffusion of strange carriers, we can refer to the heuristic description of Fig. \ref{fig:sketch}(c). A finite magnetic field introduces a lengthscale $\lambda_B=\sqrt{\hbar/eB}$; this  will eventually limit the possible extension of the hops during the diffusion processes, 
that in zero-field would instead diverge at low temperature as $\lambda_{dB}\sim 1/\sqrt{T}$, so that at large fields $D\sim (\lambda_{B})^2/\tau^{(Q)}\sim T/B$. 
Dimensional analysis suggests that if we associate to each of these lengths a characteristic frequency through 
$\lambda \sim \sqrt{\hbar/m\omega}$,
these should combine as $\omega_{tot}^2=\omega_{dB}^2+\omega_B^2$, hence modifying the diffusion length as  $\lambda_{tot}^{-4}=\lambda_{dB}^{-4}+\lambda_B^{-4}$. Replacing this in $D=(\lambda_{tot})^2/\tau^{(Q)}$  directly implies
\begin{equation}
    \rho^{(Q)}(T,B)=\rho^{(Q)}(T) \ \times \ \left[ 1+ A \left(\frac{B}{k_BT}\right)^2\right]^{1/2} 
    \label{eq:MR}
\end{equation}
with $A=(e\hbar/m)^2$,
which is the quadrature scaling 
of Ref. \cite{Ayres21}. We nte that being untied to the existence of a Fermi surface, this contribution is isotropic \cite{Grissonnanche2021}.

Similarly, the experimental fact that extrinsic disorder shifts the resistivity curves rigidly without affecting the slope \cite{RullierPRL03} can be understood by introducing a characteristic localization length $L$ and assuming that it combines with the de Broglie length as $\lambda_{tot}^{-2}=\lambda_{dB}^{-2}+L^{-2}$.

\paragraph{Bringing back the quantum in the optics.---}
The fact that the carrier wavefunctions extend over many classical mean-free-paths (Fig. \ref{fig:sketch}(b)) implies that the charge dynamics should exhibit memory effects beyond the independent collision approximation, invalidating the simple exponential decay of the momentum.  Such memory effects, of quantum origin, should show up in particular in the frequency-dependent conductivity, fundamentally altering the Drude shape. 

To demonstrate this, we use the Kubo formula for the optical conductivity \cite{Kubo,Kubobook} expressed in terms of the carriers' quantum diffusion in time \cite{AuerbachPRB10,Ciuchi11}. 
The latter is defined by the time-dependent quantum-mechanical spread $ \Delta X^2(\mathrm{t})\equiv \langle [ \hat X(\mathrm{t}) - \hat X(0) ]^2 \rangle$ in a given  direction, where $\hat X(\mathrm{t})=\sum_{i=1}^N \hat x_i(\mathrm{t})$
is  the total position operator of $N$ electrons and $\langle \cdots \rangle= \mathrm{Tr}[ e^{-\beta H} (\ldots)]/Z$ denotes the quantum thermodynamic average (we use here $\mathrm{t}$ for time, not to be confused with the transfer integral $t$). 
The real part of the optical conductivity can be written exactly as \cite{AuerbachPRB10,Ciuchi11}
\begin{equation}
   \label{eq:relation}
\mathrm{Re} \ \sigma(\omega)=-\frac{e^2\omega^2}{\nu}\frac{\tanh 
(\hbar\omega/2k_BT)}{\hbar\omega} \ \mathrm{Re}
   \int_0^\infty e^{i\omega \mathrm{t}} \Delta X^2(\mathrm{t})  d\mathrm{t}
\end{equation}
with  $e$ the electric charge and $\nu$ the volume.
The prefactor in Eq. (\ref{eq:relation}) is crucial for the  discussion that follows. 
It entails the non-commutation between position and momentum, and
it is therefore an explicit manifestation of the uncertainty principle arising when $\hbar \neq 0$.

Inverting Eq. (\ref{eq:relation}) \cite{Ciuchi11} and assuming the Drude form for $Re \ \sigma(\omega)$  yields, in the classical limit $\hbar=0$, a spread that is initially ballistic, $\Delta X^2(\mathrm{t}) \propto v^2 \mathrm{t}^2$, then diffusive $\Delta X^2(\mathrm{t}) \propto D \ \mathrm{t}$ after a time $\tau$ \cite{supplement}. Let us assume the same general form of $\Delta X^2(\mathrm{t})$ to hold in the quantum case, i.e. we consider a motion that is first ballistic then diffusive, as has been inferred from numerical data in the strongly correlated limit \cite{supplement}.
The only formal change induced by $\hbar$ will then be in the prefactor: the latter goes from the classical $1/(2k_BT)$ for $\hbar=0$, to the full quantum expression $\tanh 
(\hbar\omega/2k_BT)/\hbar\omega$ and $D\to D^{(Q)}$ at finite $\hbar$,  leading to
\begin{equation}
    \mathrm{Re} \ \sigma^{(Q)}(\omega)=\frac{\tanh 
(\hbar\omega/2k_BT)}{\hbar\omega/2k_BT} \frac{\sigma^{(Q)}}{1+ (\omega \tau)^2}.
    \label{eq:DrudeQ}
\end{equation}
This result highlights the fact that the same Planckian 
time $\tau^{(Q)}=\hbar/k_BT$ that determines the d.c. limit Eq. (\ref{eq:Planckianrho}), also controls 
the frequency-dependent response.

\begin{figure}
    \centering
    \includegraphics[width=8.5cm]{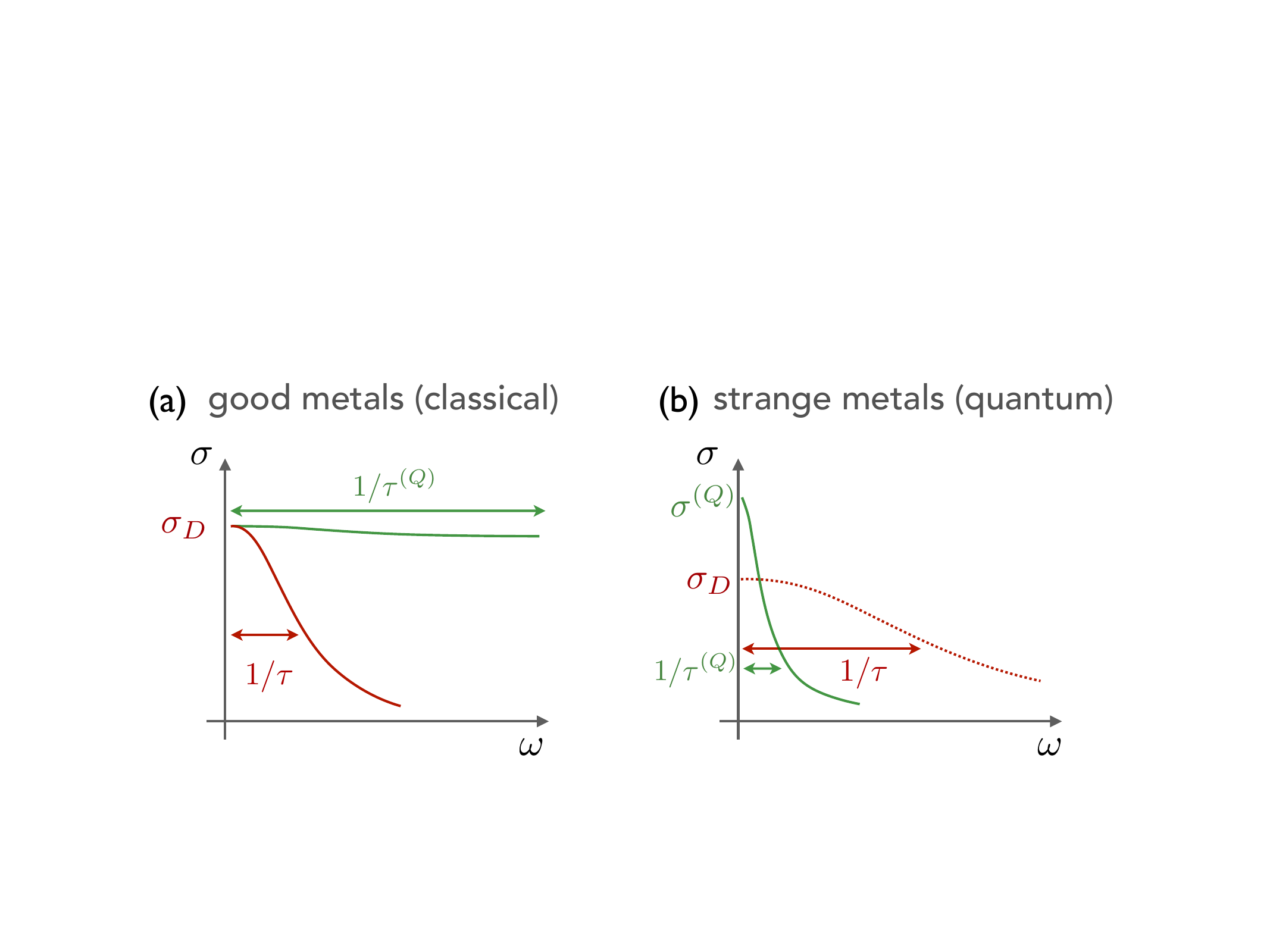}
    \caption{\textbf{Quantum Drude formula.} (a) In good metals,  $\hbar/\tau \ll k_BT$, the prefactor in Eq. (\ref{eq:DrudeQ}) is $\approx 1$ (green) and  the Drude result is recovered with $\sigma^{(Q)} = \sigma_D$ (red). (b) In strange metals, where $ k_BT \ll \hbar/\tau$, $\sigma^{(Q)} \gg \sigma_D$ and the optical absorption is entirely determined by the prefactor, with a power-law decay $1/\omega$ and a width set   by the Planckian scale $1/\tau^{(Q)}$, cf. Eq. (\ref{eq:DrudeQapp}).
    }
    \label{fig:DrudeQ}
\end{figure}

In good metals where by definition $\hbar/\tau \ll k_BT$, the prefactor is $\simeq 1$ across the entire spectrum and Eq. (\ref{eq:DrudeQ}) is indistinguishable from the classical Drude form (Fig. \ref{fig:DrudeQ}(a)). 
In the opposite case of strange metals, instead, the shape of the optical absorption  becomes explicitly dependent on $\hbar$.
Letting $\hbar/\tau \gg k_BT$ yields
\begin{equation}
    \mathrm{Re} \ \sigma^{(Q)}(\omega)\simeq \sigma^{(Q)} \  \frac{\tanh (\hbar \omega/2k_BT)}{\hbar \omega/2k_BT}.
    \label{eq:DrudeQapp}
\end{equation}
Eq. (\ref{eq:DrudeQapp}) is a scaling function of $\omega/T$ and  decays as $1/\omega$ instead of $1/\omega^2$, i.e.  it describes a stretched Drude peak as observed experimentally. 
The peak width is now set by the Planckian rate $1/\tau^{(Q)}$, that acts as an infrared cutoff, and it is therefore much narrower than $1/\tau$, that would be expected from  classical Drude theory (Fig. \ref{fig:DrudeQ}(b)). Yet, the reflectivity maintains its bad metal character, lacking a marked plasma edge as ubiquitously observed in experiments \cite{supplement}.

Fig. \ref{fig:sketch} (e) reports the full optical conductivity Eq. (\ref{eq:DrudeQ}) together with its imaginary part and  absolute value obtained via Kramers-Kronig transformation, at a temperature $k_BT/t=0.02$ and taking the MIR limit $\hbar/\tau=2t$.
The observed power-law decay of the real part, $1/\omega$, gives rise to $1/\omega^{2/3}$ for the absolute value, matching the exponent measured in Ref. \cite{vdMarelNature03} (see also Ref.\cite{PhillipsScience22}).

\begin{figure}
    \centering
    \includegraphics[width=8.5cm]{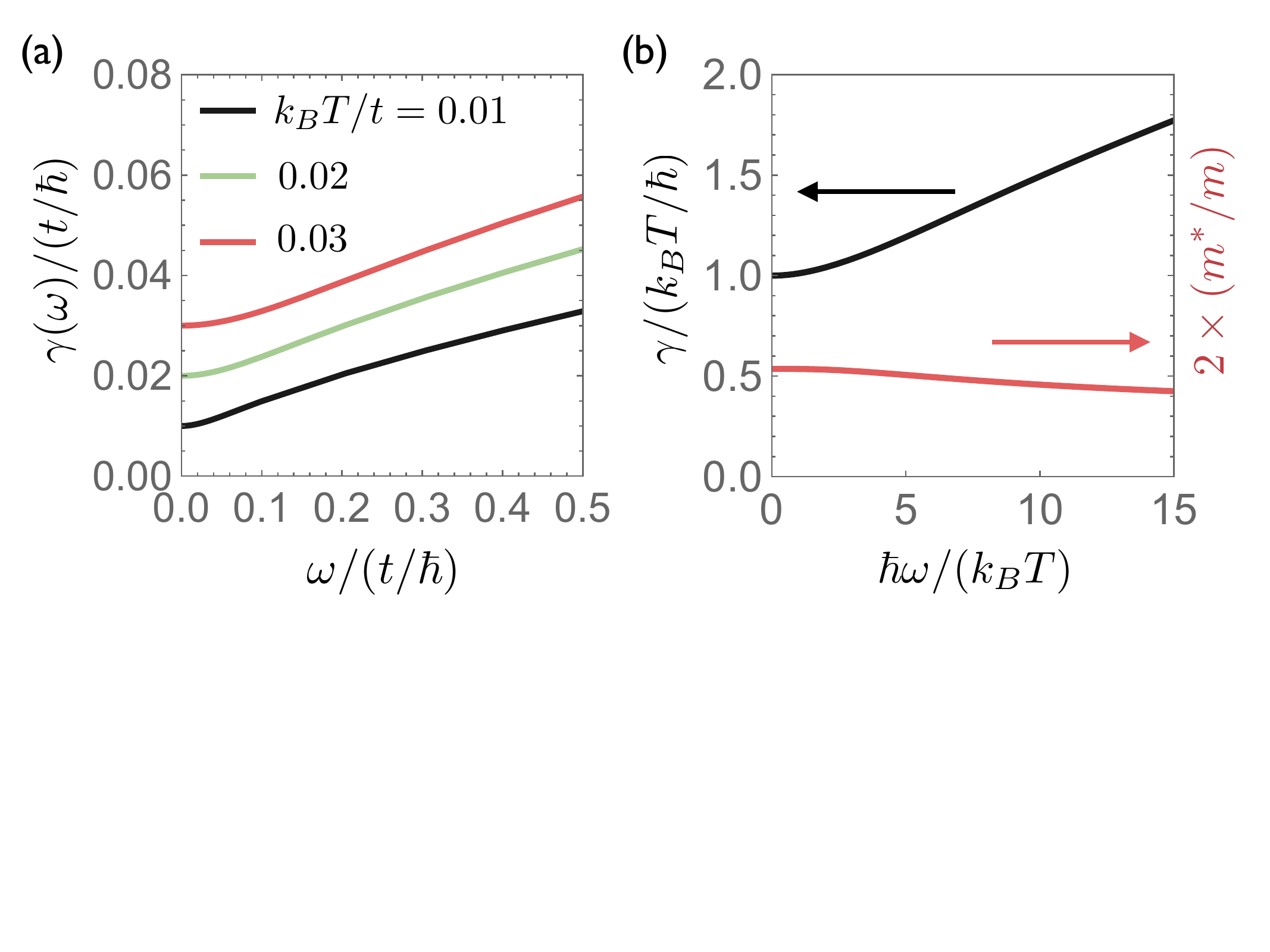}
    \caption{
    \textbf{Extended Drude analysis.} (a) Optical scattering rate as a function of frequency at different temperatures as obtained from Eq. (\ref{eq:DrudeQapp}) and (b) the same quantity rescaled in Planckian units, shown together with the optical effective mass.  }
    \label{fig:extDrude}
\end{figure}

\paragraph{Extended Drude analysis.---}
Optical conductivity spectra are customarily analyzed using the so called extended Drude model, where non-trivial memory effects 
are introduced via a frequency-dependent scattering rate $\gamma(\omega)$ and effective mass $m^*(\omega)$ as
$\sigma(\omega)=(\omega_p^2/4\pi)/[\gamma(\omega)-i  \omega \ m^*(\omega)/m]$. 
Experimentally it is found that the optical scattering rate $\gamma(\omega)$ is approximately linear in frequency, with an offset that increases with temperature. The experimental results at different temperatures show scaling behavior when plotted in units of $k_BT/\hbar$. The effective mass $m^*(\omega)/m$ is generally a slowly decreasing function of $\omega$ 
\cite{QuijadaPRB99,MichonNatComm23,BasovRMP}.
Figs. \ref{fig:extDrude}(a,b) demonstrate that Eq. (\ref{eq:DrudeQapp}) is able to rationalize all these observations, including $\omega/T$-scaling, without necessarily invoking quantum criticality \cite{VarmaRMP20,Chubukov06,Planckian-tJ}.

\begin{figure}
    \centering
    \includegraphics[width=7.5cm]{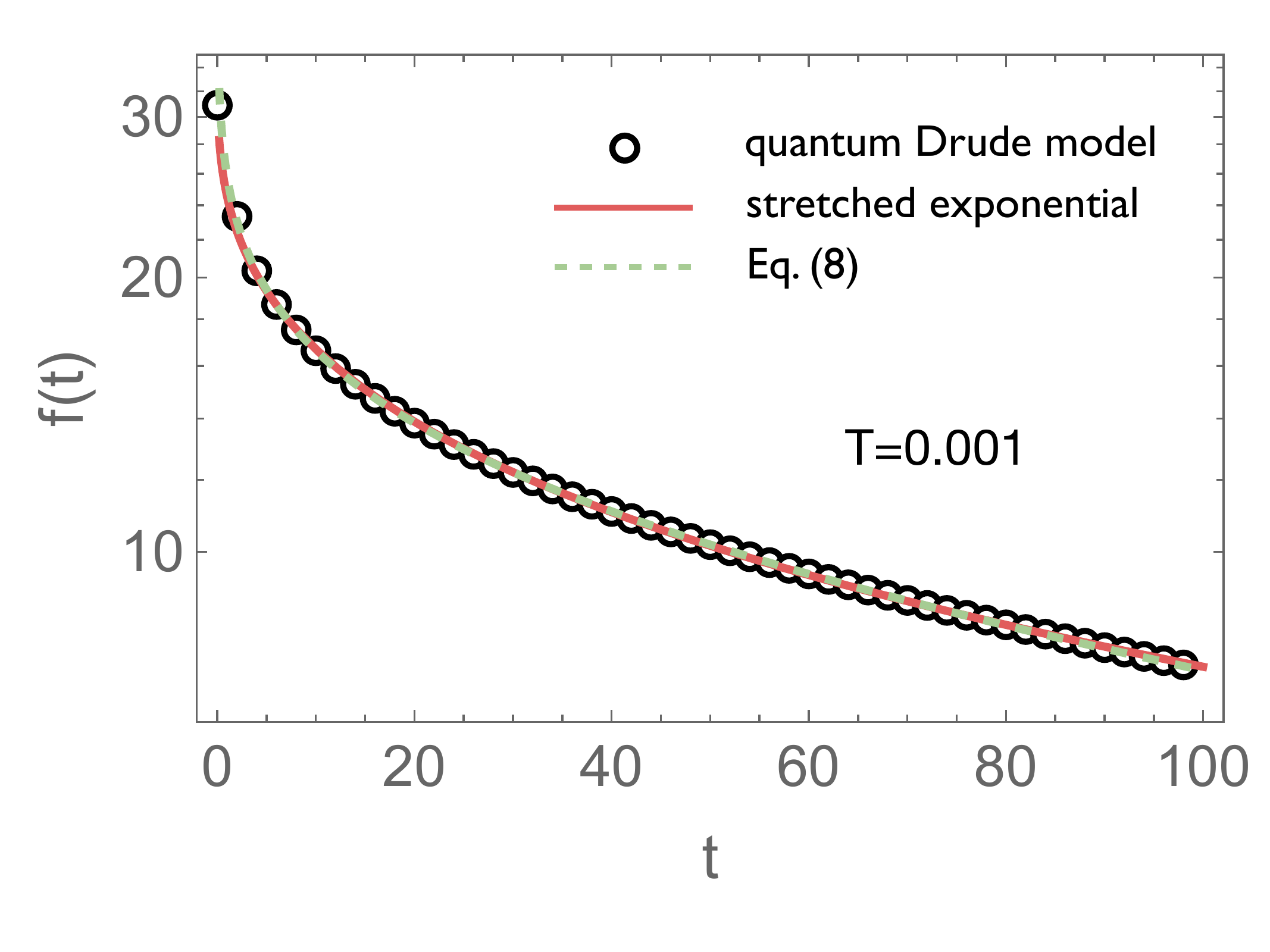}
    \caption{
    \textbf{Relaxation in the time domain.} The numerical Fourier transform of Eq. (\ref{eq:DrudeQ}) for $k_BT/t=0.001$  is shown  together with the  the analytical expression Eq. (\ref{eq:mystretch}) and the KWW stretched exponential function with  stretching exponent $b=0.32$.   All three exhibit identical non-exponential decay in the time window $\hbar/t \lesssim \mathrm{t} \lesssim \tau^{(Q)}$.
    }
    \label{fig:foft}
\end{figure}

\paragraph{Stretched exponential relaxation in the time domain.---}
Having demonstrated how memory effects manifest in the 
frequency domain, we can now ask ourselves to which type of (non-exponential) time relaxation this corresponds.
To this aim, let us take the Fourier transform of Eq. (\ref{eq:DrudeQ}), $f(\mathrm{t})=\int_0^\infty \cos (\omega \mathrm{t}) \ \sigma^{(Q)}(\omega)\ d\omega$, which  is plotted in Fig. \ref{fig:foft}. When $k_BT\ll 1/\tau$  we can use Eq. (\ref{eq:DrudeQapp}) and obtain,  at times $\mathrm{t}\lesssim 2 \tau^{(Q)}$: 
\begin{equation}
    f(\mathrm{t})
    \simeq  \frac{4\mathrm{t}\bar\sigma}{\hbar} \left\lbrack-\gamma_{EM} - \log(0.88 \ \mathrm{t} /\tau^{(Q)})\right\rbrack
    \label{eq:mystretch}
\end{equation}
with $\gamma_{EM}$  the Euler-Mascheroni constant.

As shown in Fig. \ref{fig:foft} this function is indistinguishable from the empirical Kohlrausch–Williams–Watts (KWW) stretched exponential, 
$f(\mathrm{t})= A \  e^{-(\mathrm{t}/\mathrm{t}_0)^b}$, in a wide range of $\mathrm{t}$.
The quantum Drude formula Eq. (\ref{eq:DrudeQ}) or, for dielectrics, an equivalent quantum Debye formula, may provide a physical justification to this widely observed behavior. Here the physical origin of the $\log$ is the 
Planckian time cutoff embedded in the uncertainty principle \cite{noteCD}.

\paragraph{Concluding remarks.---}
The present theory identifies strange metal behavior as a general feature arising when, due to the presence of strong scattering,  quantum uncertainty cannot be neglected in the charge diffusion mechanism. The theory is otherwise agnostic about the microscopic nature of the current carriers, that only enters through their effective mass in Eq. (\ref{eq:Dq}) and the resulting resistivity Eq. (\ref{eq:Planckianrho}). 

From a theoretical standpoint, it has been shown that weakly doped Mott-Hubbard insulators in the limit of strong Coulomb interactions $U/t\to \infty$ strictly comply with the present predictions, with the minimum diffusivity Eq. (\ref{eq:Dq}) arising from the scattering of electrons or holes by the disordered spin background \cite{Planckian-tJ}. For finite values of $U/t$,  magnetic correlations driven by the antiferromagnetic exchange $J$ as well as kinetic frustration can promote more complex emergent excitations such as Trugman loops \cite{TrugmanPRB88,DemlerPRX18}, anti-ferromagnetic polarons \cite{Batista25}  or other fractionalized excitations \cite{Coleman89,SenthilPRL03}. Thermal melting of these heavier carriers at $k_BT > J$ might explain, via Eq. (\ref{eq:Planckianrho}), the existence of two separate strange metal regimes with different slopes (i.e., different masses) observed in the calculated resistivities \cite{Kokalj,DengPRL13,Planckian-tJ}.

Experimentally, prototypical examples of strange metals include the superconducting cuprates \cite{PhillipsScience22,Bruin,Legros} as well as twisted bilayer graphene \cite{CaoPRL20,Jaoui2022} and twisted bilayer dichalcogenides \cite{XiaMakShan25}. 
In all these strongly interacting systems the scattering by spin fluctuations \cite{campbell2024strangemetalspinfluctuations,Planckian-tJ},  quantum critical modes \cite{Grilli23} or other abundant excitations \cite{AydinPNAS23}
can naturally lead to a suppression of the diffusivity to the value Eq. \ref{eq:Dq} required for the present theory to apply. Once this condition is met, the theory is compatible with the observation of strange behavior across broad doping ranges.
In this respect, several experimental probes in the cuprates, including photoemission \cite{ZXshenRMP03}, magneto-transport \cite{Ayres21,Grissonnanche2021} and optical absorption  \cite{QuijadaPRB99,vanHeumen22} suggest that strange incoherent carriers persist all the way to the strongly overdoped regime.
Understanding how strange carriers fade out upon doping, 
being progressively replaced by normal electrons and possibly coexisting with them  remains an open question.

\acknowledgments
This work was supported by the French Agence Nationale de la Recherche (ANR) under reference ANR-25-CE30-2817 (Strangemetal).
I am grateful to Sergio Ciuchi for fundamental insights on this problem.
Special thanks go to Chisa Hotta for her kind hospitality and the vivid discussions we had at the Komaba campus, where the  idea underlying this work first appeared. I also acknowledge invaluable input from K. Behnia, P. M. Caminero, A. Chubukov, P. Coleman, M. Dressel,
A. G. Grushin, E. van Heumen, N. Hussey, C. Iorio-Duval, M.-H. Julien, D. LeBoeuf, L. Mangeolle, D. Maslov, L. de' Medici, A. J. Millis and L. Rademaker.

\bibliography{minimal} 

\begin{thebibliography}{44}%
\makeatletter
\providecommand \@ifxundefined [1]{%
 \@ifx{#1\undefined}
}%
\providecommand \@ifnum [1]{%
 \ifnum #1\expandafter \@firstoftwo
 \else \expandafter \@secondoftwo
 \fi
}%
\providecommand \@ifx [1]{%
 \ifx #1\expandafter \@firstoftwo
 \else \expandafter \@secondoftwo
 \fi
}%
\providecommand \natexlab [1]{#1}%
\providecommand \enquote  [1]{``#1''}%
\providecommand \bibnamefont  [1]{#1}%
\providecommand \bibfnamefont [1]{#1}%
\providecommand \citenamefont [1]{#1}%
\providecommand \href@noop [0]{\@secondoftwo}%
\providecommand \href [0]{\begingroup \@sanitize@url \@href}%
\providecommand \@href[1]{\@@startlink{#1}\@@href}%
\providecommand \@@href[1]{\endgroup#1\@@endlink}%
\providecommand \@sanitize@url [0]{\catcode `\\12\catcode `\$12\catcode `\&12\catcode `\#12\catcode `\^12\catcode `\_12\catcode `\%12\relax}%
\providecommand \@@startlink[1]{}%
\providecommand \@@endlink[0]{}%
\providecommand \url  [0]{\begingroup\@sanitize@url \@url }%
\providecommand \@url [1]{\endgroup\@href {#1}{\urlprefix }}%
\providecommand \urlprefix  [0]{URL }%
\providecommand \Eprint [0]{\href }%
\providecommand \doibase [0]{https://doi.org/}%
\providecommand \selectlanguage [0]{\@gobble}%
\providecommand \bibinfo  [0]{\@secondoftwo}%
\providecommand \bibfield  [0]{\@secondoftwo}%
\providecommand \translation [1]{[#1]}%
\providecommand \BibitemOpen [0]{}%
\providecommand \bibitemStop [0]{}%
\providecommand \bibitemNoStop [0]{.\EOS\space}%
\providecommand \EOS [0]{\spacefactor3000\relax}%
\providecommand \BibitemShut  [1]{\csname bibitem#1\endcsname}%
\let\auto@bib@innerbib\@empty
\bibitem [{\citenamefont {Phillips}\ \emph {et~al.}(2022)\citenamefont {Phillips}, \citenamefont {Hussey},\ and\ \citenamefont {Abbamonte}}]{PhillipsScience22}%
  \BibitemOpen
  \bibfield  {author} {\bibinfo {author} {\bibfnamefont {P.~W.}\ \bibnamefont {Phillips}}, \bibinfo {author} {\bibfnamefont {N.~E.}\ \bibnamefont {Hussey}},\ and\ \bibinfo {author} {\bibfnamefont {P.}~\bibnamefont {Abbamonte}},\ }\bibfield  {title} {\bibinfo {title} {Stranger than metals},\ }\href {https://doi.org/10.1126/science.abh4273} {\bibfield  {journal} {\bibinfo  {journal} {Science}\ }\textbf {\bibinfo {volume} {377}},\ \bibinfo {pages} {eabh4273} (\bibinfo {year} {2022})}\BibitemShut {NoStop}%
\bibitem [{\citenamefont {Bruin}\ \emph {et~al.}(2013)\citenamefont {Bruin}, \citenamefont {Sakai}, \citenamefont {Perry},\ and\ \citenamefont {Mackenzie}}]{Bruin}%
  \BibitemOpen
  \bibfield  {author} {\bibinfo {author} {\bibfnamefont {J.~A.~N.}\ \bibnamefont {Bruin}}, \bibinfo {author} {\bibfnamefont {H.}~\bibnamefont {Sakai}}, \bibinfo {author} {\bibfnamefont {R.~S.}\ \bibnamefont {Perry}},\ and\ \bibinfo {author} {\bibfnamefont {A.~P.}\ \bibnamefont {Mackenzie}},\ }\bibfield  {title} {\bibinfo {title} {Similarity of scattering rates in metals showing {T}-linear resistivity},\ }\href {https://doi.org/10.1126/science.1227612} {\bibfield  {journal} {\bibinfo  {journal} {Science}\ }\textbf {\bibinfo {volume} {339}},\ \bibinfo {pages} {804} (\bibinfo {year} {2013})}\BibitemShut {NoStop}%
\bibitem [{\citenamefont {Legros}\ \emph {et~al.}(2019)\citenamefont {Legros}, \citenamefont {Benhabib}, \citenamefont {Tabis}, \citenamefont {Lalibert{\'e}}, \citenamefont {Dion}, \citenamefont {Lizaire}, \citenamefont {Vignolle}, \citenamefont {Vignolles}, \citenamefont {Raffy}, \citenamefont {Li}, \citenamefont {{Auban-Senzier}}, \citenamefont {{Doiron-Leyraud}}, \citenamefont {Fournier}, \citenamefont {Colson}, \citenamefont {Taillefer},\ and\ \citenamefont {Proust}}]{Legros}%
  \BibitemOpen
  \bibfield  {author} {\bibinfo {author} {\bibfnamefont {A.}~\bibnamefont {Legros}}, \bibinfo {author} {\bibfnamefont {S.}~\bibnamefont {Benhabib}}, \bibinfo {author} {\bibfnamefont {W.}~\bibnamefont {Tabis}}, \bibinfo {author} {\bibfnamefont {F.}~\bibnamefont {Lalibert{\'e}}}, \bibinfo {author} {\bibfnamefont {M.}~\bibnamefont {Dion}}, \bibinfo {author} {\bibfnamefont {M.}~\bibnamefont {Lizaire}}, \bibinfo {author} {\bibfnamefont {B.}~\bibnamefont {Vignolle}}, \bibinfo {author} {\bibfnamefont {D.}~\bibnamefont {Vignolles}}, \bibinfo {author} {\bibfnamefont {H.}~\bibnamefont {Raffy}}, \bibinfo {author} {\bibfnamefont {Z.~Z.}\ \bibnamefont {Li}}, \bibinfo {author} {\bibfnamefont {P.}~\bibnamefont {{Auban-Senzier}}}, \bibinfo {author} {\bibfnamefont {N.}~\bibnamefont {{Doiron-Leyraud}}}, \bibinfo {author} {\bibfnamefont {P.}~\bibnamefont {Fournier}}, \bibinfo {author} {\bibfnamefont {D.}~\bibnamefont {Colson}}, \bibinfo {author} {\bibfnamefont {L.}~\bibnamefont {Taillefer}},\ and\ \bibinfo {author}
  {\bibfnamefont {C.}~\bibnamefont {Proust}},\ }\bibfield  {title} {\bibinfo {title} {Universal {{T-linear}} resistivity and {{Planckian}} dissipation in overdoped cuprates},\ }\href {https://doi.org/10.1038/s41567-018-0334-2} {\bibfield  {journal} {\bibinfo  {journal} {Nature Physics}\ }\textbf {\bibinfo {volume} {15}},\ \bibinfo {pages} {142} (\bibinfo {year} {2019})}\BibitemShut {NoStop}%
\bibitem [{\citenamefont {van~der Marel}\ \emph {et~al.}(2003)\citenamefont {van~der Marel}, \citenamefont {Molegraaf}, \citenamefont {Zaanen}, \citenamefont {Nussinov}, \citenamefont {Carbone}, \citenamefont {Damascelli}, \citenamefont {Eisaki}, \citenamefont {Greven}, \citenamefont {Kes},\ and\ \citenamefont {Li}}]{vdMarelNature03}%
  \BibitemOpen
  \bibfield  {author} {\bibinfo {author} {\bibfnamefont {D.}~\bibnamefont {van~der Marel}}, \bibinfo {author} {\bibfnamefont {H.~J.~A.}\ \bibnamefont {Molegraaf}}, \bibinfo {author} {\bibfnamefont {J.}~\bibnamefont {Zaanen}}, \bibinfo {author} {\bibfnamefont {Z.}~\bibnamefont {Nussinov}}, \bibinfo {author} {\bibfnamefont {F.}~\bibnamefont {Carbone}}, \bibinfo {author} {\bibfnamefont {A.}~\bibnamefont {Damascelli}}, \bibinfo {author} {\bibfnamefont {H.}~\bibnamefont {Eisaki}}, \bibinfo {author} {\bibfnamefont {M.}~\bibnamefont {Greven}}, \bibinfo {author} {\bibfnamefont {P.~H.}\ \bibnamefont {Kes}},\ and\ \bibinfo {author} {\bibfnamefont {M.}~\bibnamefont {Li}},\ }\bibfield  {title} {\bibinfo {title} {Quantum critical behaviour in a high-{{Tc}} superconductor},\ }\href {https://doi.org/10.1038/nature01978} {\bibfield  {journal} {\bibinfo  {journal} {Nature}\ }\textbf {\bibinfo {volume} {425}},\ \bibinfo {pages} {271} (\bibinfo {year} {2003})}\BibitemShut {NoStop}%
\bibitem [{\citenamefont {Michon}\ \emph {et~al.}(2023)\citenamefont {Michon}, \citenamefont {Berthod}, \citenamefont {Rischau}, \citenamefont {Ataei}, \citenamefont {Chen}, \citenamefont {Komiya}, \citenamefont {Ono}, \citenamefont {Taillefer}, \citenamefont {{van der Marel}},\ and\ \citenamefont {Georges}}]{MichonNatComm23}%
  \BibitemOpen
  \bibfield  {author} {\bibinfo {author} {\bibfnamefont {B.}~\bibnamefont {Michon}}, \bibinfo {author} {\bibfnamefont {C.}~\bibnamefont {Berthod}}, \bibinfo {author} {\bibfnamefont {C.~W.}\ \bibnamefont {Rischau}}, \bibinfo {author} {\bibfnamefont {A.}~\bibnamefont {Ataei}}, \bibinfo {author} {\bibfnamefont {L.}~\bibnamefont {Chen}}, \bibinfo {author} {\bibfnamefont {S.}~\bibnamefont {Komiya}}, \bibinfo {author} {\bibfnamefont {S.}~\bibnamefont {Ono}}, \bibinfo {author} {\bibfnamefont {L.}~\bibnamefont {Taillefer}}, \bibinfo {author} {\bibfnamefont {D.}~\bibnamefont {{van der Marel}}},\ and\ \bibinfo {author} {\bibfnamefont {A.}~\bibnamefont {Georges}},\ }\bibfield  {title} {\bibinfo {title} {Reconciling scaling of the optical conductivity of cuprate superconductors with {{Planckian}} resistivity and specific heat},\ }\href {https://doi.org/10.1038/s41467-023-38762-5} {\bibfield  {journal} {\bibinfo  {journal} {Nature Communications}\ }\textbf {\bibinfo {volume} {14}},\ \bibinfo {pages} {3033} (\bibinfo
  {year} {2023})}\BibitemShut {NoStop}%
\bibitem [{\citenamefont {van Heumen}\ \emph {et~al.}(2022)\citenamefont {van Heumen}, \citenamefont {Feng}, \citenamefont {Cassanelli}, \citenamefont {Neubrand}, \citenamefont {de~Jager}, \citenamefont {Berben}, \citenamefont {Huang}, \citenamefont {Kondo}, \citenamefont {Takeuchi},\ and\ \citenamefont {Zaanen}}]{vanHeumen22}%
  \BibitemOpen
  \bibfield  {author} {\bibinfo {author} {\bibfnamefont {E.}~\bibnamefont {van Heumen}}, \bibinfo {author} {\bibfnamefont {X.}~\bibnamefont {Feng}}, \bibinfo {author} {\bibfnamefont {S.}~\bibnamefont {Cassanelli}}, \bibinfo {author} {\bibfnamefont {L.}~\bibnamefont {Neubrand}}, \bibinfo {author} {\bibfnamefont {L.}~\bibnamefont {de~Jager}}, \bibinfo {author} {\bibfnamefont {M.}~\bibnamefont {Berben}}, \bibinfo {author} {\bibfnamefont {Y.}~\bibnamefont {Huang}}, \bibinfo {author} {\bibfnamefont {T.}~\bibnamefont {Kondo}}, \bibinfo {author} {\bibfnamefont {T.}~\bibnamefont {Takeuchi}},\ and\ \bibinfo {author} {\bibfnamefont {J.}~\bibnamefont {Zaanen}},\ }\bibfield  {title} {\bibinfo {title} {Strange metal electrodynamics across the phase diagram of {${\mathrm{Bi}}_{2\ensuremath{-}x}{\mathrm{Pb}}_{x}{\mathrm{Sr}}_{2\ensuremath{-}y}{\mathrm{La}}_{y}{\mathrm{CuO}}_{6+\ensuremath{\delta}}$} cuprates},\ }\href {https://doi.org/10.1103/PhysRevB.106.054515} {\bibfield  {journal} {\bibinfo  {journal} {Phys. Rev. B}\
  }\textbf {\bibinfo {volume} {106}},\ \bibinfo {pages} {054515} (\bibinfo {year} {2022})}\BibitemShut {NoStop}%
\bibitem [{\citenamefont {Greenwood}(1958)}]{Greenwood58}%
  \BibitemOpen
  \bibfield  {author} {\bibinfo {author} {\bibfnamefont {D.~A.}\ \bibnamefont {Greenwood}},\ }\bibfield  {title} {\bibinfo {title} {The {Boltzmann} equation in the theory of electrical conduction in metals},\ }\href {https://doi.org/10.1088/0370-1328/71/4/306} {\bibfield  {journal} {\bibinfo  {journal} {Proceedings of the Physical Society}\ }\textbf {\bibinfo {volume} {71}},\ \bibinfo {pages} {585} (\bibinfo {year} {1958})}\BibitemShut {NoStop}%
\bibitem [{\citenamefont {Ashcroft}\ and\ \citenamefont {Mermin}(1976)}]{Ashcroft}%
  \BibitemOpen
  \bibfield  {author} {\bibinfo {author} {\bibfnamefont {N.~W.}\ \bibnamefont {Ashcroft}}\ and\ \bibinfo {author} {\bibfnamefont {N.~D.}\ \bibnamefont {Mermin}},\ }\href@noop {} {\emph {\bibinfo {title} {{Solid state physics}}}}\ (\bibinfo  {publisher} {Holt, Rinehart and Winston},\ \bibinfo {address} {New York, NY},\ \bibinfo {year} {1976})\BibitemShut {NoStop}%
\bibitem [{\citenamefont {Fratini}\ \emph {et~al.}(2025)\citenamefont {Fratini}, \citenamefont {Duchemin}, \citenamefont {Ralko},\ and\ \citenamefont {Ciuchi}}]{Planckian-tJ}%
  \BibitemOpen
  \bibfield  {author} {\bibinfo {author} {\bibfnamefont {S.}~\bibnamefont {Fratini}}, \bibinfo {author} {\bibfnamefont {I.}~\bibnamefont {Duchemin}}, \bibinfo {author} {\bibfnamefont {A.}~\bibnamefont {Ralko}},\ and\ \bibinfo {author} {\bibfnamefont {S.}~\bibnamefont {Ciuchi}},\ }\href {https://arxiv.org/abs/2412.04322} {\bibinfo {title} {Strange metal transport from coupling to fluctuating spins}} (\bibinfo {year} {2025}),\ \Eprint {https://arxiv.org/abs/2412.04322} {arXiv:2412.04322 [cond-mat.str-el]} \BibitemShut {NoStop}%
\bibitem [{not()}]{noteCD}%
  \BibitemOpen
  \href@noop {} {\ }\bibinfo {note} {In the same way, the frequency-dependence Eq. (\ref{eq:DrudeQ}) is very similar to the Cole-Davidson function $\sigma(\omega)=\sigma_0/(1-i\omega \tau)^c$ that is commonly used in the context of dielectric relaxation \cite{Hill78}, and that seems to reproduce both the theoretical \cite{Planckian-tJ} and experimental \cite{vanHeumen22} optical absorption spectra of strange metals (Fig. \ref{fig:sketch} (e), dashed lines, $c=0.78$).}\BibitemShut {Stop}%
\bibitem [{\citenamefont {Sommer}\ \emph {et~al.}(2011)\citenamefont {Sommer}, \citenamefont {Ku}, \citenamefont {Roati},\ and\ \citenamefont {Zwierlein}}]{Sommer11}%
  \BibitemOpen
  \bibfield  {author} {\bibinfo {author} {\bibfnamefont {A.}~\bibnamefont {Sommer}}, \bibinfo {author} {\bibfnamefont {M.}~\bibnamefont {Ku}}, \bibinfo {author} {\bibfnamefont {G.}~\bibnamefont {Roati}},\ and\ \bibinfo {author} {\bibfnamefont {M.~W.}\ \bibnamefont {Zwierlein}},\ }\bibfield  {title} {\bibinfo {title} {Universal spin transport in a strongly interacting {{Fermi}} gas},\ }\href {https://doi.org/10.1038/nature09989} {\bibfield  {journal} {\bibinfo  {journal} {Nature}\ }\textbf {\bibinfo {volume} {472}},\ \bibinfo {pages} {201} (\bibinfo {year} {2011})}\BibitemShut {NoStop}%
\bibitem [{\citenamefont {Bardon}\ \emph {et~al.}(2014)\citenamefont {Bardon}, \citenamefont {Beattie}, \citenamefont {Luciuk}, \citenamefont {Cairncross}, \citenamefont {Fine}, \citenamefont {Cheng}, \citenamefont {Edge}, \citenamefont {Taylor}, \citenamefont {Zhang}, \citenamefont {Trotzky},\ and\ \citenamefont {Thywissen}}]{Bardon14}%
  \BibitemOpen
  \bibfield  {author} {\bibinfo {author} {\bibfnamefont {A.~B.}\ \bibnamefont {Bardon}}, \bibinfo {author} {\bibfnamefont {S.}~\bibnamefont {Beattie}}, \bibinfo {author} {\bibfnamefont {C.}~\bibnamefont {Luciuk}}, \bibinfo {author} {\bibfnamefont {W.}~\bibnamefont {Cairncross}}, \bibinfo {author} {\bibfnamefont {D.}~\bibnamefont {Fine}}, \bibinfo {author} {\bibfnamefont {N.~S.}\ \bibnamefont {Cheng}}, \bibinfo {author} {\bibfnamefont {G.~J.~A.}\ \bibnamefont {Edge}}, \bibinfo {author} {\bibfnamefont {E.}~\bibnamefont {Taylor}}, \bibinfo {author} {\bibfnamefont {S.}~\bibnamefont {Zhang}}, \bibinfo {author} {\bibfnamefont {S.}~\bibnamefont {Trotzky}},\ and\ \bibinfo {author} {\bibfnamefont {J.~H.}\ \bibnamefont {Thywissen}},\ }\bibfield  {title} {\bibinfo {title} {Transverse demagnetization dynamics of a unitary fermi gas},\ }\href {https://doi.org/10.1126/science.1247425} {\bibfield  {journal} {\bibinfo  {journal} {Science}\ }\textbf {\bibinfo {volume} {344}},\ \bibinfo {pages} {722} (\bibinfo {year}
  {2014})}\BibitemShut {NoStop}%
\bibitem [{\citenamefont {Aydin}\ \emph {et~al.}(2024)\citenamefont {Aydin}, \citenamefont {Keski-Rahkonen},\ and\ \citenamefont {Heller}}]{AydinPNAS23}%
  \BibitemOpen
  \bibfield  {author} {\bibinfo {author} {\bibfnamefont {A.}~\bibnamefont {Aydin}}, \bibinfo {author} {\bibfnamefont {J.}~\bibnamefont {Keski-Rahkonen}},\ and\ \bibinfo {author} {\bibfnamefont {E.~J.}\ \bibnamefont {Heller}},\ }\bibfield  {title} {\bibinfo {title} {Quantum acoustics unravels {Planckian} resistivity},\ }\href {https://doi.org/10.1073/pnas.2404853121} {\bibfield  {journal} {\bibinfo  {journal} {Proceedings of the National Academy of Sciences}\ }\textbf {\bibinfo {volume} {121}},\ \bibinfo {pages} {e2404853121} (\bibinfo {year} {2024})}\BibitemShut {NoStop}%
\bibitem [{Note1()}]{Note1}%
  \BibitemOpen
  \bibinfo {note} {On a lattice, the lower bound Eq. (\ref {eq:Dq}) is equivalent to the Mott-Ioffe-Regel (MIR) condition stating that wavelike coherence is lost at each hop: setting $\ell =a$ with $\ell =v \tau $ the mean-free-path and $a$ the nearest-neighbor distance, and taking $v=2 t a/\hbar $ describing ballistic motion of electrons with transfer integral $t$ yields a scattering rate $\tau _{MIR}=\hbar /2t$ and a diffusion constant $D=2t a^2/\hbar $. Using $m= \hbar ^2/(2ta^2)$ leads to $D^{(Q)}=\hbar /m$.}\BibitemShut {Stop}%
\bibitem [{\citenamefont {Kokalj}(2017)}]{Kokalj}%
  \BibitemOpen
  \bibfield  {author} {\bibinfo {author} {\bibfnamefont {J.}~\bibnamefont {Kokalj}},\ }\bibfield  {title} {\bibinfo {title} {Bad-metallic behavior of doped {Mott} insulators},\ }\href {https://doi.org/10.1103/PhysRevB.95.041110} {\bibfield  {journal} {\bibinfo  {journal} {Phys. Rev. B}\ }\textbf {\bibinfo {volume} {95}},\ \bibinfo {pages} {041110(R)} (\bibinfo {year} {2017})}\BibitemShut {NoStop}%
\bibitem [{\citenamefont {Grilli}\ \emph {et~al.}(2023)\citenamefont {Grilli}, \citenamefont {Di~Castro}, \citenamefont {Mirarchi}, \citenamefont {Seibold},\ and\ \citenamefont {Caprara}}]{Grilli23}%
  \BibitemOpen
  \bibfield  {author} {\bibinfo {author} {\bibfnamefont {M.}~\bibnamefont {Grilli}}, \bibinfo {author} {\bibfnamefont {C.}~\bibnamefont {Di~Castro}}, \bibinfo {author} {\bibfnamefont {G.}~\bibnamefont {Mirarchi}}, \bibinfo {author} {\bibfnamefont {G.}~\bibnamefont {Seibold}},\ and\ \bibinfo {author} {\bibfnamefont {S.}~\bibnamefont {Caprara}},\ }\bibfield  {title} {\bibinfo {title} {Dissipative quantum criticality as a source of strange metal behavior},\ }\bibfield  {journal} {\bibinfo  {journal} {Symmetry}\ }\textbf {\bibinfo {volume} {15}},\ \href {https://doi.org/10.3390/sym15030569} {10.3390/sym15030569} (\bibinfo {year} {2023})\BibitemShut {NoStop}%
\bibitem [{\citenamefont {Mahan}(2000)}]{Mahan}%
  \BibitemOpen
  \bibfield  {author} {\bibinfo {author} {\bibfnamefont {G.~D.}\ \bibnamefont {Mahan}},\ }\href@noop {} {\emph {\bibinfo {title} {Many-Particle Physics}}}\ (\bibinfo  {publisher} {Kluwer Academic/Plenum Publisher, New Yok},\ \bibinfo {year} {2000})\BibitemShut {NoStop}%
\bibitem [{\citenamefont {Li}\ \emph {et~al.}(2018)\citenamefont {Li}, \citenamefont {Chen},\ and\ \citenamefont {Fisher}}]{LiFischerPRB18}%
  \BibitemOpen
  \bibfield  {author} {\bibinfo {author} {\bibfnamefont {Y.}~\bibnamefont {Li}}, \bibinfo {author} {\bibfnamefont {X.}~\bibnamefont {Chen}},\ and\ \bibinfo {author} {\bibfnamefont {M.~P.~A.}\ \bibnamefont {Fisher}},\ }\bibfield  {title} {\bibinfo {title} {Quantum zeno effect and the many-body entanglement transition},\ }\href {https://doi.org/10.1103/PhysRevB.98.205136} {\bibfield  {journal} {\bibinfo  {journal} {Phys. Rev. B}\ }\textbf {\bibinfo {volume} {98}},\ \bibinfo {pages} {205136} (\bibinfo {year} {2018})}\BibitemShut {NoStop}%
\bibitem [{\citenamefont {Ciuchi}\ \emph {et~al.}(2011)\citenamefont {Ciuchi}, \citenamefont {Fratini},\ and\ \citenamefont {Mayou}}]{Ciuchi11}%
  \BibitemOpen
  \bibfield  {author} {\bibinfo {author} {\bibfnamefont {S.}~\bibnamefont {Ciuchi}}, \bibinfo {author} {\bibfnamefont {S.}~\bibnamefont {Fratini}},\ and\ \bibinfo {author} {\bibfnamefont {D.}~\bibnamefont {Mayou}},\ }\bibfield  {title} {\bibinfo {title} {Transient localization in crystalline organic semiconductors},\ }\href {https://doi.org/10.1103/PhysRevB.83.081202} {\bibfield  {journal} {\bibinfo  {journal} {Phys. Rev. B}\ }\textbf {\bibinfo {volume} {83}},\ \bibinfo {pages} {081202(R)} (\bibinfo {year} {2011})}\BibitemShut {NoStop}%
\bibitem [{\citenamefont {Fratini}\ \emph {et~al.}(2016)\citenamefont {Fratini}, \citenamefont {Mayou},\ and\ \citenamefont {Ciuchi}}]{Fratini-AdvMat16}%
  \BibitemOpen
  \bibfield  {author} {\bibinfo {author} {\bibfnamefont {S.}~\bibnamefont {Fratini}}, \bibinfo {author} {\bibfnamefont {D.}~\bibnamefont {Mayou}},\ and\ \bibinfo {author} {\bibfnamefont {S.}~\bibnamefont {Ciuchi}},\ }\bibfield  {title} {\bibinfo {title} {The transient localization scenario for charge transport in crystalline organic materials},\ }\href {https://doi.org/10.1002/adfm.201502386} {\bibfield  {journal} {\bibinfo  {journal} {Adv. Funct. Mater}\ }\textbf {\bibinfo {volume} {26}},\ \bibinfo {pages} {2292} (\bibinfo {year} {2016})}\BibitemShut {NoStop}%
\bibitem [{\citenamefont {Giannini}\ \emph {et~al.}(2022)\citenamefont {Giannini}, \citenamefont {Peng}, \citenamefont {Cupellini}, \citenamefont {Padula}, \citenamefont {Carof},\ and\ \citenamefont {Blumberger}}]{Giannini}%
  \BibitemOpen
  \bibfield  {author} {\bibinfo {author} {\bibfnamefont {S.}~\bibnamefont {Giannini}}, \bibinfo {author} {\bibfnamefont {W.-T.}\ \bibnamefont {Peng}}, \bibinfo {author} {\bibfnamefont {L.}~\bibnamefont {Cupellini}}, \bibinfo {author} {\bibfnamefont {D.}~\bibnamefont {Padula}}, \bibinfo {author} {\bibfnamefont {A.}~\bibnamefont {Carof}},\ and\ \bibinfo {author} {\bibfnamefont {J.}~\bibnamefont {Blumberger}},\ }\bibfield  {title} {\bibinfo {title} {Exciton transport in molecular organic semiconductors boosted by transient quantum delocalization},\ }\bibfield  {journal} {\bibinfo  {journal} {Nature Communications}\ }\textbf {\bibinfo {volume} {13}},\ \href {https://doi.org/10.1038/s41467-022-30308-5} {10.1038/s41467-022-30308-5} (\bibinfo {year} {2022})\BibitemShut {NoStop}%
\bibitem [{\citenamefont {Ayres}\ \emph {et~al.}(2021)\citenamefont {Ayres}, \citenamefont {Berben}, \citenamefont {{\v C}ulo}, \citenamefont {Hsu}, \citenamefont {Van~Heumen}, \citenamefont {Huang}, \citenamefont {Zaanen}, \citenamefont {Kondo}, \citenamefont {Takeuchi}, \citenamefont {Cooper}, \citenamefont {Putzke}, \citenamefont {Friedemann}, \citenamefont {Carrington},\ and\ \citenamefont {Hussey}}]{Ayres21}%
  \BibitemOpen
  \bibfield  {author} {\bibinfo {author} {\bibfnamefont {J.}~\bibnamefont {Ayres}}, \bibinfo {author} {\bibfnamefont {M.}~\bibnamefont {Berben}}, \bibinfo {author} {\bibfnamefont {M.}~\bibnamefont {{\v C}ulo}}, \bibinfo {author} {\bibfnamefont {Y.-T.}\ \bibnamefont {Hsu}}, \bibinfo {author} {\bibfnamefont {E.}~\bibnamefont {Van~Heumen}}, \bibinfo {author} {\bibfnamefont {Y.}~\bibnamefont {Huang}}, \bibinfo {author} {\bibfnamefont {J.}~\bibnamefont {Zaanen}}, \bibinfo {author} {\bibfnamefont {T.}~\bibnamefont {Kondo}}, \bibinfo {author} {\bibfnamefont {T.}~\bibnamefont {Takeuchi}}, \bibinfo {author} {\bibfnamefont {J.~R.}\ \bibnamefont {Cooper}}, \bibinfo {author} {\bibfnamefont {C.}~\bibnamefont {Putzke}}, \bibinfo {author} {\bibfnamefont {S.}~\bibnamefont {Friedemann}}, \bibinfo {author} {\bibfnamefont {A.}~\bibnamefont {Carrington}},\ and\ \bibinfo {author} {\bibfnamefont {N.~E.}\ \bibnamefont {Hussey}},\ }\bibfield  {title} {\bibinfo {title} {Incoherent transport across the strange-metal regime of
  overdoped cuprates},\ }\href {https://doi.org/10.1038/s41586-021-03622-z} {\bibfield  {journal} {\bibinfo  {journal} {Nature}\ }\textbf {\bibinfo {volume} {595}},\ \bibinfo {pages} {661} (\bibinfo {year} {2021})}\BibitemShut {NoStop}%
\bibitem [{\citenamefont {Grissonnanche}\ \emph {et~al.}(2021)\citenamefont {Grissonnanche}, \citenamefont {Fang}, \citenamefont {Legros}, \citenamefont {Verret}, \citenamefont {Lalibert{\' e}}, \citenamefont {Collignon}, \citenamefont {Zhou}, \citenamefont {Graf}, \citenamefont {Goddard}, \citenamefont {Taillefer},\ and\ \citenamefont {Ramshaw}}]{Grissonnanche2021}%
  \BibitemOpen
  \bibfield  {author} {\bibinfo {author} {\bibfnamefont {G.}~\bibnamefont {Grissonnanche}}, \bibinfo {author} {\bibfnamefont {Y.}~\bibnamefont {Fang}}, \bibinfo {author} {\bibfnamefont {A.}~\bibnamefont {Legros}}, \bibinfo {author} {\bibfnamefont {S.}~\bibnamefont {Verret}}, \bibinfo {author} {\bibfnamefont {F.}~\bibnamefont {Lalibert{\' e}}}, \bibinfo {author} {\bibfnamefont {C.}~\bibnamefont {Collignon}}, \bibinfo {author} {\bibfnamefont {J.}~\bibnamefont {Zhou}}, \bibinfo {author} {\bibfnamefont {D.}~\bibnamefont {Graf}}, \bibinfo {author} {\bibfnamefont {P.~A.}\ \bibnamefont {Goddard}}, \bibinfo {author} {\bibfnamefont {L.}~\bibnamefont {Taillefer}},\ and\ \bibinfo {author} {\bibfnamefont {B.~J.}\ \bibnamefont {Ramshaw}},\ }\bibfield  {title} {\bibinfo {title} {Linear-in temperature resistivity from an isotropic {Planckian} scattering rate},\ }\href@noop {} {\bibfield  {journal} {\bibinfo  {journal} {Nature}\ }\textbf {\bibinfo {volume} {595}},\ \bibinfo {pages} {667} (\bibinfo {year} {2021})}\BibitemShut
  {NoStop}%
\bibitem [{\citenamefont {Rullier-Albenque}\ \emph {et~al.}(2003)\citenamefont {Rullier-Albenque}, \citenamefont {Alloul},\ and\ \citenamefont {Tourbot}}]{RullierPRL03}%
  \BibitemOpen
  \bibfield  {author} {\bibinfo {author} {\bibfnamefont {F.}~\bibnamefont {Rullier-Albenque}}, \bibinfo {author} {\bibfnamefont {H.}~\bibnamefont {Alloul}},\ and\ \bibinfo {author} {\bibfnamefont {R.}~\bibnamefont {Tourbot}},\ }\bibfield  {title} {\bibinfo {title} {Influence of pair breaking and phase fluctuations on disordered high-${T}_{c}$ cuprate superconductors},\ }\href@noop {} {\bibfield  {journal} {\bibinfo  {journal} {Phys. Rev. Lett.}\ }\textbf {\bibinfo {volume} {91}},\ \bibinfo {pages} {047001} (\bibinfo {year} {2003})}\BibitemShut {NoStop}%
\bibitem [{\citenamefont {Kubo}(1957)}]{Kubo}%
  \BibitemOpen
  \bibfield  {author} {\bibinfo {author} {\bibfnamefont {R.}~\bibnamefont {Kubo}},\ }\bibfield  {title} {\bibinfo {title} {Statistical-mechanical theory of irreversible processes. {I}. {General} theory and simple applications to magnetic and conduction problems},\ }\href {https://doi.org/10.1143/JPSJ.12.570} {\bibfield  {journal} {\bibinfo  {journal} {Journal of the Physical Society of Japan}\ }\textbf {\bibinfo {volume} {12}},\ \bibinfo {pages} {570} (\bibinfo {year} {1957})}\BibitemShut {NoStop}%
\bibitem [{\citenamefont {Ryogo~Kubo}(1991)}]{Kubobook}%
  \BibitemOpen
  \bibfield  {author} {\bibinfo {author} {\bibfnamefont {N.~H.}\ \bibnamefont {Ryogo~Kubo}, \bibfnamefont {Morikazu~Toda}},\ }\href@noop {} {\emph {\bibinfo {title} {Statistical Physics {II}: Nonequilibrium Statistical Mechanics}}}\ (\bibinfo  {publisher} {{Springer}},\ \bibinfo {address} {{New York}},\ \bibinfo {year} {1991})\BibitemShut {NoStop}%
\bibitem [{\citenamefont {Lindner}\ and\ \citenamefont {Auerbach}(2010)}]{AuerbachPRB10}%
  \BibitemOpen
  \bibfield  {author} {\bibinfo {author} {\bibfnamefont {N.~H.}\ \bibnamefont {Lindner}}\ and\ \bibinfo {author} {\bibfnamefont {A.}~\bibnamefont {Auerbach}},\ }\bibfield  {title} {\bibinfo {title} {Conductivity of hard core bosons: A paradigm of a bad metal},\ }\href {https://doi.org/10.1103/PhysRevB.81.054512} {\bibfield  {journal} {\bibinfo  {journal} {Phys. Rev. B}\ }\textbf {\bibinfo {volume} {81}},\ \bibinfo {pages} {054512} (\bibinfo {year} {2010})}\BibitemShut {NoStop}%
\bibitem [{sup()}]{supplement}%
  \BibitemOpen
  \href@noop {} {\ }\bibinfo {note} {See Supplemental Material at [URL] for details on the derivation of Eq. (6), including a representative numerical example and an illustration of the emergence of displaced Drude peaks in the case of subdiffusive motion.}\BibitemShut {Stop}%
\bibitem [{\citenamefont {Quijada}\ \emph {et~al.}(1999)\citenamefont {Quijada}, \citenamefont {Tanner}, \citenamefont {Kelley}, \citenamefont {Onellion}, \citenamefont {Berger},\ and\ \citenamefont {Margaritondo}}]{QuijadaPRB99}%
  \BibitemOpen
  \bibfield  {author} {\bibinfo {author} {\bibfnamefont {M.~A.}\ \bibnamefont {Quijada}}, \bibinfo {author} {\bibfnamefont {D.~B.}\ \bibnamefont {Tanner}}, \bibinfo {author} {\bibfnamefont {R.~J.}\ \bibnamefont {Kelley}}, \bibinfo {author} {\bibfnamefont {M.}~\bibnamefont {Onellion}}, \bibinfo {author} {\bibfnamefont {H.}~\bibnamefont {Berger}},\ and\ \bibinfo {author} {\bibfnamefont {G.}~\bibnamefont {Margaritondo}},\ }\bibfield  {title} {\bibinfo {title} {Anisotropy in the ab-plane optical properties of {Bi}$_2${Sr}$_{2}${CaCu}$_{2}${O}$_{8}$ single-domain crystals},\ }\href {https://doi.org/10.1103/PhysRevB.60.14917} {\bibfield  {journal} {\bibinfo  {journal} {Phys. Rev. B}\ }\textbf {\bibinfo {volume} {60}},\ \bibinfo {pages} {14917} (\bibinfo {year} {1999})}\BibitemShut {NoStop}%
\bibitem [{\citenamefont {Basov}\ and\ \citenamefont {Timusk}(2005)}]{BasovRMP}%
  \BibitemOpen
  \bibfield  {author} {\bibinfo {author} {\bibfnamefont {D.~N.}\ \bibnamefont {Basov}}\ and\ \bibinfo {author} {\bibfnamefont {T.}~\bibnamefont {Timusk}},\ }\bibfield  {title} {\bibinfo {title} {Electrodynamics of high-${T}_{c}$ superconductors},\ }\href {https://doi.org/10.1103/RevModPhys.77.721} {\bibfield  {journal} {\bibinfo  {journal} {Rev. Mod. Phys.}\ }\textbf {\bibinfo {volume} {77}},\ \bibinfo {pages} {721} (\bibinfo {year} {2005})}\BibitemShut {NoStop}%
\bibitem [{\citenamefont {Varma}(2020)}]{VarmaRMP20}%
  \BibitemOpen
  \bibfield  {author} {\bibinfo {author} {\bibfnamefont {C.~M.}\ \bibnamefont {Varma}},\ }\bibfield  {title} {\bibinfo {title} {Colloquium: Linear in temperature resistivity and associated mysteries including high temperature superconductivity},\ }\href {https://doi.org/10.1103/RevModPhys.92.031001} {\bibfield  {journal} {\bibinfo  {journal} {Rev. Mod. Phys.}\ }\textbf {\bibinfo {volume} {92}},\ \bibinfo {pages} {031001} (\bibinfo {year} {2020})}\BibitemShut {NoStop}%
\bibitem [{\citenamefont {Norman}\ and\ \citenamefont {Chubukov}(2006)}]{Chubukov06}%
  \BibitemOpen
  \bibfield  {author} {\bibinfo {author} {\bibfnamefont {M.~R.}\ \bibnamefont {Norman}}\ and\ \bibinfo {author} {\bibfnamefont {A.~V.}\ \bibnamefont {Chubukov}},\ }\bibfield  {title} {\bibinfo {title} {High-frequency behavior of the infrared conductivity of cuprates},\ }\href {https://doi.org/10.1103/PhysRevB.73.140501} {\bibfield  {journal} {\bibinfo  {journal} {Phys. Rev. B}\ }\textbf {\bibinfo {volume} {73}},\ \bibinfo {pages} {140501} (\bibinfo {year} {2006})}\BibitemShut {NoStop}%
\bibitem [{\citenamefont {Trugman}(1988)}]{TrugmanPRB88}%
  \BibitemOpen
  \bibfield  {author} {\bibinfo {author} {\bibfnamefont {S.~A.}\ \bibnamefont {Trugman}},\ }\bibfield  {title} {\bibinfo {title} {Interaction of holes in a hubbard antiferromagnet and high-temperature superconductivity},\ }\href {https://doi.org/10.1103/PhysRevB.37.1597} {\bibfield  {journal} {\bibinfo  {journal} {Phys. Rev. B}\ }\textbf {\bibinfo {volume} {37}},\ \bibinfo {pages} {1597} (\bibinfo {year} {1988})}\BibitemShut {NoStop}%
\bibitem [{\citenamefont {Grusdt}\ \emph {et~al.}(2018)\citenamefont {Grusdt}, \citenamefont {K\'anasz-Nagy}, \citenamefont {Bohrdt}, \citenamefont {Chiu}, \citenamefont {Ji}, \citenamefont {Greiner}, \citenamefont {Greif},\ and\ \citenamefont {Demler}}]{DemlerPRX18}%
  \BibitemOpen
  \bibfield  {author} {\bibinfo {author} {\bibfnamefont {F.}~\bibnamefont {Grusdt}}, \bibinfo {author} {\bibfnamefont {M.}~\bibnamefont {K\'anasz-Nagy}}, \bibinfo {author} {\bibfnamefont {A.}~\bibnamefont {Bohrdt}}, \bibinfo {author} {\bibfnamefont {C.~S.}\ \bibnamefont {Chiu}}, \bibinfo {author} {\bibfnamefont {G.}~\bibnamefont {Ji}}, \bibinfo {author} {\bibfnamefont {M.}~\bibnamefont {Greiner}}, \bibinfo {author} {\bibfnamefont {D.}~\bibnamefont {Greif}},\ and\ \bibinfo {author} {\bibfnamefont {E.}~\bibnamefont {Demler}},\ }\bibfield  {title} {\bibinfo {title} {Parton theory of magnetic polarons: Mesonic resonances and signatures in dynamics},\ }\href {https://doi.org/10.1103/PhysRevX.8.011046} {\bibfield  {journal} {\bibinfo  {journal} {Phys. Rev. X}\ }\textbf {\bibinfo {volume} {8}},\ \bibinfo {pages} {011046} (\bibinfo {year} {2018})}\BibitemShut {NoStop}%
\bibitem [{\citenamefont {Zhang}\ \emph {et~al.}(2025)\citenamefont {Zhang}, \citenamefont {Batista},\ and\ \citenamefont {Zhang}}]{Batista25}%
  \BibitemOpen
  \bibfield  {author} {\bibinfo {author} {\bibfnamefont {Y.}~\bibnamefont {Zhang}}, \bibinfo {author} {\bibfnamefont {C.}~\bibnamefont {Batista}},\ and\ \bibinfo {author} {\bibfnamefont {Y.}~\bibnamefont {Zhang}},\ }\href {https://arxiv.org/abs/2506.16464} {\bibinfo {title} {Antiferromagnetism and tightly bound {Cooper} pairs induced by kinetic frustration}} (\bibinfo {year} {2025}),\ \Eprint {https://arxiv.org/abs/2506.16464} {arXiv:2506.16464 [cond-mat.str-el]} \BibitemShut {NoStop}%
\bibitem [{\citenamefont {Coleman}\ and\ \citenamefont {Andrei}(1989)}]{Coleman89}%
  \BibitemOpen
  \bibfield  {author} {\bibinfo {author} {\bibfnamefont {P.}~\bibnamefont {Coleman}}\ and\ \bibinfo {author} {\bibfnamefont {N.}~\bibnamefont {Andrei}},\ }\bibfield  {title} {\bibinfo {title} {Kondo-stabilised spin liquids and heavy fermion superconductivity},\ }\href {https://doi.org/10.1088/0953-8984/1/26/003} {\bibfield  {journal} {\bibinfo  {journal} {Journal of Physics: Condensed Matter}\ }\textbf {\bibinfo {volume} {1}},\ \bibinfo {pages} {4057} (\bibinfo {year} {1989})}\BibitemShut {NoStop}%
\bibitem [{\citenamefont {Senthil}\ \emph {et~al.}(2003)\citenamefont {Senthil}, \citenamefont {Sachdev},\ and\ \citenamefont {Vojta}}]{SenthilPRL03}%
  \BibitemOpen
  \bibfield  {author} {\bibinfo {author} {\bibfnamefont {T.}~\bibnamefont {Senthil}}, \bibinfo {author} {\bibfnamefont {S.}~\bibnamefont {Sachdev}},\ and\ \bibinfo {author} {\bibfnamefont {M.}~\bibnamefont {Vojta}},\ }\bibfield  {title} {\bibinfo {title} {Fractionalized {Fermi} liquids},\ }\href {https://doi.org/10.1103/PhysRevLett.90.216403} {\bibfield  {journal} {\bibinfo  {journal} {Phys. Rev. Lett.}\ }\textbf {\bibinfo {volume} {90}},\ \bibinfo {pages} {216403} (\bibinfo {year} {2003})}\BibitemShut {NoStop}%
\bibitem [{\citenamefont {Deng}\ \emph {et~al.}(2013)\citenamefont {Deng}, \citenamefont {Mravlje}, \citenamefont {\ifmmode~\check{Z}\else \v{Z}\fi{}itko}, \citenamefont {Ferrero}, \citenamefont {Kotliar},\ and\ \citenamefont {Georges}}]{DengPRL13}%
  \BibitemOpen
  \bibfield  {author} {\bibinfo {author} {\bibfnamefont {X.}~\bibnamefont {Deng}}, \bibinfo {author} {\bibfnamefont {J.}~\bibnamefont {Mravlje}}, \bibinfo {author} {\bibfnamefont {R.}~\bibnamefont {\ifmmode~\check{Z}\else \v{Z}\fi{}itko}}, \bibinfo {author} {\bibfnamefont {M.}~\bibnamefont {Ferrero}}, \bibinfo {author} {\bibfnamefont {G.}~\bibnamefont {Kotliar}},\ and\ \bibinfo {author} {\bibfnamefont {A.}~\bibnamefont {Georges}},\ }\bibfield  {title} {\bibinfo {title} {How bad metals turn good: Spectroscopic signatures of resilient quasiparticles},\ }\href {https://doi.org/10.1103/PhysRevLett.110.086401} {\bibfield  {journal} {\bibinfo  {journal} {Phys. Rev. Lett.}\ }\textbf {\bibinfo {volume} {110}},\ \bibinfo {pages} {086401} (\bibinfo {year} {2013})}\BibitemShut {NoStop}%
\bibitem [{\citenamefont {Cao}\ \emph {et~al.}(2020)\citenamefont {Cao}, \citenamefont {Chowdhury}, \citenamefont {Rodan-Legrain}, \citenamefont {Rubies-Bigorda}, \citenamefont {Watanabe}, \citenamefont {Taniguchi}, \citenamefont {Senthil},\ and\ \citenamefont {Jarillo-Herrero}}]{CaoPRL20}%
  \BibitemOpen
  \bibfield  {author} {\bibinfo {author} {\bibfnamefont {Y.}~\bibnamefont {Cao}}, \bibinfo {author} {\bibfnamefont {D.}~\bibnamefont {Chowdhury}}, \bibinfo {author} {\bibfnamefont {D.}~\bibnamefont {Rodan-Legrain}}, \bibinfo {author} {\bibfnamefont {O.}~\bibnamefont {Rubies-Bigorda}}, \bibinfo {author} {\bibfnamefont {K.}~\bibnamefont {Watanabe}}, \bibinfo {author} {\bibfnamefont {T.}~\bibnamefont {Taniguchi}}, \bibinfo {author} {\bibfnamefont {T.}~\bibnamefont {Senthil}},\ and\ \bibinfo {author} {\bibfnamefont {P.}~\bibnamefont {Jarillo-Herrero}},\ }\bibfield  {title} {\bibinfo {title} {Strange metal in magic-angle graphene with near {Planckian} dissipation},\ }\href {https://doi.org/10.1103/PhysRevLett.124.076801} {\bibfield  {journal} {\bibinfo  {journal} {Phys. Rev. Lett.}\ }\textbf {\bibinfo {volume} {124}},\ \bibinfo {pages} {076801} (\bibinfo {year} {2020})}\BibitemShut {NoStop}%
\bibitem [{\citenamefont {Jaoui}\ \emph {et~al.}(2022)\citenamefont {Jaoui}, \citenamefont {Das}, \citenamefont {Di~Battista}, \citenamefont {D{\'\i}ez-M{\'e}rida}, \citenamefont {Lu}, \citenamefont {Watanabe}, \citenamefont {Taniguchi}, \citenamefont {Ishizuka}, \citenamefont {Levitov},\ and\ \citenamefont {Efetov}}]{Jaoui2022}%
  \BibitemOpen
  \bibfield  {author} {\bibinfo {author} {\bibfnamefont {A.}~\bibnamefont {Jaoui}}, \bibinfo {author} {\bibfnamefont {I.}~\bibnamefont {Das}}, \bibinfo {author} {\bibfnamefont {G.}~\bibnamefont {Di~Battista}}, \bibinfo {author} {\bibfnamefont {J.}~\bibnamefont {D{\'\i}ez-M{\'e}rida}}, \bibinfo {author} {\bibfnamefont {X.}~\bibnamefont {Lu}}, \bibinfo {author} {\bibfnamefont {K.}~\bibnamefont {Watanabe}}, \bibinfo {author} {\bibfnamefont {T.}~\bibnamefont {Taniguchi}}, \bibinfo {author} {\bibfnamefont {H.}~\bibnamefont {Ishizuka}}, \bibinfo {author} {\bibfnamefont {L.}~\bibnamefont {Levitov}},\ and\ \bibinfo {author} {\bibfnamefont {D.~K.}\ \bibnamefont {Efetov}},\ }\bibfield  {title} {\bibinfo {title} {Quantum critical behaviour in magic-angle twisted bilayer graphene},\ }\href@noop {} {\bibfield  {journal} {\bibinfo  {journal} {Nat. Phys.}\ }\textbf {\bibinfo {volume} {18}},\ \bibinfo {pages} {633} (\bibinfo {year} {2022})}\BibitemShut {NoStop}%
\bibitem [{\citenamefont {Xia}\ \emph {et~al.}(2025)\citenamefont {Xia}, \citenamefont {Han}, \citenamefont {Zhu}, \citenamefont {Zhang}, \citenamefont {Knüppel}, \citenamefont {Watanabe}, \citenamefont {Taniguchi}, \citenamefont {Mak},\ and\ \citenamefont {Shan}}]{XiaMakShan25}%
  \BibitemOpen
  \bibfield  {author} {\bibinfo {author} {\bibfnamefont {Y.}~\bibnamefont {Xia}}, \bibinfo {author} {\bibfnamefont {Z.}~\bibnamefont {Han}}, \bibinfo {author} {\bibfnamefont {J.}~\bibnamefont {Zhu}}, \bibinfo {author} {\bibfnamefont {Y.}~\bibnamefont {Zhang}}, \bibinfo {author} {\bibfnamefont {P.}~\bibnamefont {Knüppel}}, \bibinfo {author} {\bibfnamefont {K.}~\bibnamefont {Watanabe}}, \bibinfo {author} {\bibfnamefont {T.}~\bibnamefont {Taniguchi}}, \bibinfo {author} {\bibfnamefont {K.~F.}\ \bibnamefont {Mak}},\ and\ \bibinfo {author} {\bibfnamefont {J.}~\bibnamefont {Shan}},\ }\href {https://arxiv.org/abs/2508.02662} {\bibinfo {title} {Simulating high-temperature superconductivity in {Moir\'e} {WSe$_2$}}} (\bibinfo {year} {2025}),\ \Eprint {https://arxiv.org/abs/2508.02662} {arXiv:2508.02662 [cond-mat.supr-con]} \BibitemShut {NoStop}%
\bibitem [{\citenamefont {Campbell}\ \emph {et~al.}(2025)\citenamefont {Campbell}, \citenamefont {Frachet}, \citenamefont {Oliviero}, \citenamefont {Kurosawa}, \citenamefont {Momono}, \citenamefont {Oda}, \citenamefont {Chang}, \citenamefont {Vignolles}, \citenamefont {Proust},\ and\ \citenamefont {LeBoeuf}}]{campbell2024strangemetalspinfluctuations}%
  \BibitemOpen
  \bibfield  {author} {\bibinfo {author} {\bibfnamefont {D.~J.}\ \bibnamefont {Campbell}}, \bibinfo {author} {\bibfnamefont {M.}~\bibnamefont {Frachet}}, \bibinfo {author} {\bibfnamefont {V.}~\bibnamefont {Oliviero}}, \bibinfo {author} {\bibfnamefont {T.}~\bibnamefont {Kurosawa}}, \bibinfo {author} {\bibfnamefont {N.}~\bibnamefont {Momono}}, \bibinfo {author} {\bibfnamefont {M.}~\bibnamefont {Oda}}, \bibinfo {author} {\bibfnamefont {J.}~\bibnamefont {Chang}}, \bibinfo {author} {\bibfnamefont {D.}~\bibnamefont {Vignolles}}, \bibinfo {author} {\bibfnamefont {C.}~\bibnamefont {Proust}},\ and\ \bibinfo {author} {\bibfnamefont {D.}~\bibnamefont {LeBoeuf}},\ }\bibfield  {title} {\bibinfo {title} {Impact of low-energy spin fluctuations on the strange metal in a cuprate superconductor},\ }\href@noop {} {\bibfield  {journal} {\bibinfo  {journal} {Nat. Phys.}\ }\textbf {\bibinfo {volume} {21}},\ \bibinfo {pages} {1759} (\bibinfo {year} {2025})}\BibitemShut {NoStop}%
\bibitem [{\citenamefont {Damascelli}\ \emph {et~al.}(2003)\citenamefont {Damascelli}, \citenamefont {Hussain},\ and\ \citenamefont {Shen}}]{ZXshenRMP03}%
  \BibitemOpen
  \bibfield  {author} {\bibinfo {author} {\bibfnamefont {A.}~\bibnamefont {Damascelli}}, \bibinfo {author} {\bibfnamefont {Z.}~\bibnamefont {Hussain}},\ and\ \bibinfo {author} {\bibfnamefont {Z.-X.}\ \bibnamefont {Shen}},\ }\bibfield  {title} {\bibinfo {title} {Angle-resolved photoemission studies of the cuprate superconductors},\ }\href {https://doi.org/10.1103/RevModPhys.75.473} {\bibfield  {journal} {\bibinfo  {journal} {Rev. Mod. Phys.}\ }\textbf {\bibinfo {volume} {75}},\ \bibinfo {pages} {473} (\bibinfo {year} {2003})}\BibitemShut {NoStop}%
\bibitem [{\citenamefont {Hill}(1978)}]{Hill78}%
  \BibitemOpen
  \bibfield  {author} {\bibinfo {author} {\bibfnamefont {R.~M.}\ \bibnamefont {Hill}},\ }\bibfield  {title} {\bibinfo {title} {Characterisation of dielectric loss in solids and liquids},\ }\href {https://doi.org/10.1038/275096a0} {\bibfield  {journal} {\bibinfo  {journal} {Nature}\ }\textbf {\bibinfo {volume} {275}},\ \bibinfo {pages} {96} (\bibinfo {year} {1978})}\BibitemShut {NoStop}%
\end{thebibliography}%

\onecolumngrid

\pagebreak
\newpage

\renewcommand{\thefigure}{S\arabic{figure}}
\renewcommand{\thetable}{S\arabic{table}}
\renewcommand{\theequation}{S\arabic{equation}}
\renewcommand{\thepage}{S\arabic{page}}
\setcounter{figure}{0}
\setcounter{table}{0}
\setcounter{equation}{0}
\setcounter{page}{1}

\section*{
Supplementary Materials for \\
Minimal Theory of Strange Carriers}

\section{Derivation of Eq. (6)}

To demonstrate Eq. (6) we first invert Eq. (5), obtaining [19]: 
\begin{equation}
   \label{eq:relinv}
  \Delta X^2(\mathrm{t})=\frac{2\hbar \nu}{\pi e^2}
  \int_0^\infty (1- \cos \omega \mathrm{t})\frac{ Re \
 \sigma(\omega)}{\omega\tanh 
(\hbar\omega/2k_BT)}
    d\omega.
\end{equation}
Assuming the Drude form for $Re \ \sigma(\omega)$ and performing the integral in the classical limit $\hbar=0$ yields the following function 
\begin{equation}
   \label{eq:relinvclas}
  \Delta X^2(\mathrm{t})_{Drude}=\frac{4 \nu k_BT}{\pi e^2}
  \int_0^\infty \frac{(1- \cos \omega \mathrm{t})}{\omega^2}\frac{ \sigma}{1+(\omega \tau)^2}
    d\omega.
\end{equation}
To obtain the quantum Drude formula  Eq. (6) we assume that the general shape of $\Delta X^2(\mathrm{t})_{Drude}$ is unchanged (see Section \ref{sec:num} for a numerical justification). We can now apply Eq. (5) to the result, restoring a finite $\hbar$ to account for quantum uncertainty: 
\begin{equation}
   \label{eq:relation}
\mathrm{Re} \ \sigma(\omega)=-\frac{e^2\omega^2}{\nu}\frac{\tanh 
(\hbar\omega/2k_BT)}{\hbar\omega} \ \mathrm{Re}
   \int_0^\infty e^{i\omega \mathrm{t}} \Delta X^2(\mathrm{t})_{Drude}  d\mathrm{t},
\end{equation}
Substituting Eq. (\ref{eq:relinvclas}) into Eq. (\ref{eq:relation}), using the  diffusion constant Eq. (1) for the quantum case  (i.e. replacing the classical d.c. value $\sigma$ with $\sigma^{(Q)}$) and performing the double integral yields, for all $\omega>0$, 
\begin{eqnarray}
\nonumber \mathrm{Re} \ \sigma(\omega)& = & -\frac{e^2\omega^2}{\nu}\frac{\tanh 
(\hbar\omega/2k_BT)}{\hbar\omega}  \frac{4 \nu k_BT}{\pi e^2} \ \mathrm{Re}
   \int_0^\infty d\mathrm{t} \cos{\omega \mathrm{t}} 
  \int_0^\infty  d\omega^\prime \frac{(1- \cos \omega^\prime \mathrm{t})}{\omega^{\prime \ 2}}\frac{ \sigma^{(Q)}}{1+(\omega^\prime \tau)^2} \\
  & = & \frac{\tanh 
(\hbar\omega/2k_BT)}{\hbar\omega/2k_BT} \frac{ \sigma^{(Q)}}{1+(\omega \tau)^2}   ,
   \label{eq:demonstration}
\end{eqnarray}
which is Eq. (6) of the main text.

\section{Quantum spread in the many-body correlated state}
\label{sec:num}

\begin{figure}
    \centering
    \includegraphics[width=16cm]{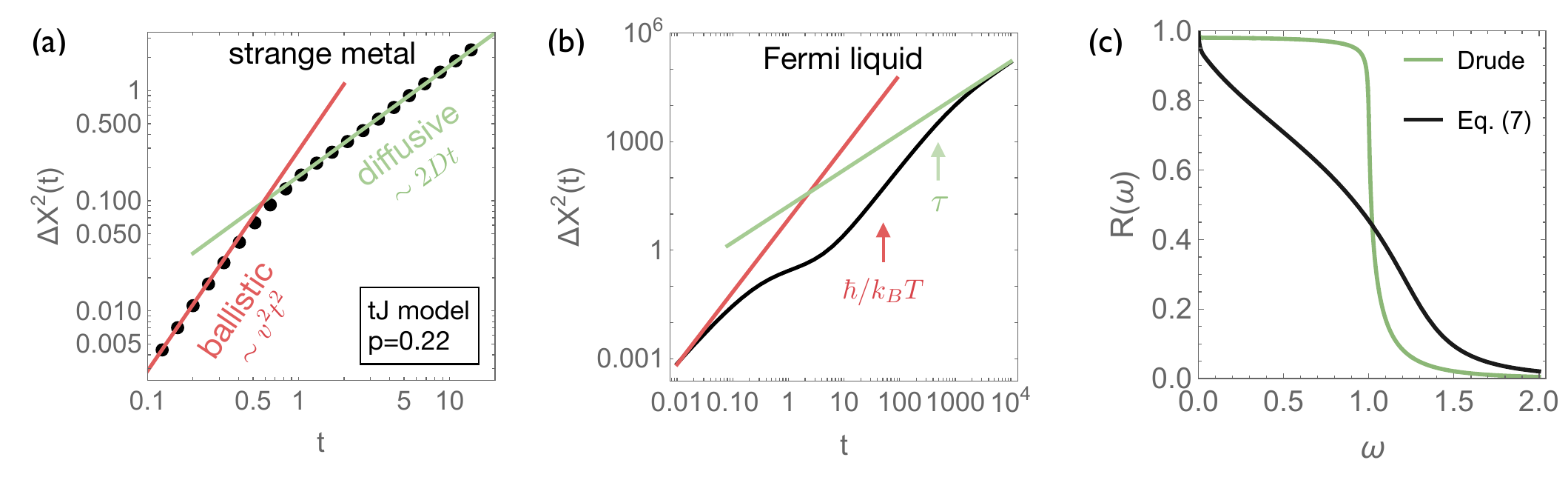}
    \caption{(a) Many-body quantum spread calculated in the t-J model at doping $p=0.22$ and temperature $T/t=0.4$, well within the strange metal regime (see text). Time is in units of the inverse of the transfer integral $\hbar/t$. (b) Same quantity for a Fermi liquid, assuming a scattering rate $\hbar/\tau=0.002 t$ and a much higher $k_BT=10 \hbar/\tau$. (c) Reflectivity derived from  the quantum Drude formula Eq. (7) (black) compared with the classical Drude reflectivity for the same  peak width given by the Planckian scale $k_BT/\hbar=0.01$ (green). Frequencies are in units of the plasma frequency $\omega_P=1$.}
    \label{fig:DeltaXnum}
\end{figure}

Fig. \ref{fig:DeltaXnum}(a) reports a representative plot of the many-body quantum spread $\Delta X^2(\mathrm{t})$ obtained through Eq. (\ref{eq:relinv}) upon integrating the  optical conductivity of the t-J model on the square lattice, that was calculated in Ref.  [9] with exact numerical accuracy through the finite temperature Lanczos method. The chosen parameters are doping $p=0.22$ and temperature $T/t=0.4$, well within the strange metal regime (cf. Fig. 4A of that work) for $J/t=0.4$, $J^\prime=J/2$. As suggested by this plot, the fact that $\Delta X^2(\mathrm{t})$ is formally analogous to the classical result would  appear to be an identifying feature of strange metals.
The value of the many-body diffusion constant $D$ 
is close to the quantum diffusivity unit $\hbar/m$ (see Fig. 3C in Ref. [9]).

We also report in Fig. \ref{fig:DeltaXnum}(b) for comparison the many-body quantum spread $\Delta X^2(\mathrm{t})$ corresponding to a Fermi liquid, calculated by performing the integral Eq. (\ref{eq:relinv}) on the analytical formulas of Ref. [45], in the regime of $\hbar/\tau\ll k_BT$ where the theory applies.


\section{Subdiffusive motion and displaced Drude peaks}

We briefly consider the possibility that the carrier motion is subdiffusive in a given time window, before normal diffusion is established at long times. This idea of "transient subdiffusion"  is similar to the one underlying transient localization theory, 
and that has been shown to give rise to displaced Drude peaks [19,20]. The latter originate from the fact that for subdiffusive motion the conductivity strictly vanishes when $\omega\to 0$.  Because the  conductivity also decays to zero when $\omega\to \infty$, subdiffusion implies the existence of a maximum at finite frequency.
If the subdiffusive behavior  stops at a time $\tau_{in}$, a finite
d.c. conductivity  $\simeq \sigma(\omega= 1/\tau_{in})$ is restored
(see Refs. [19,20] for details).

\begin{figure}
    \centering
    \includegraphics[width=10cm]{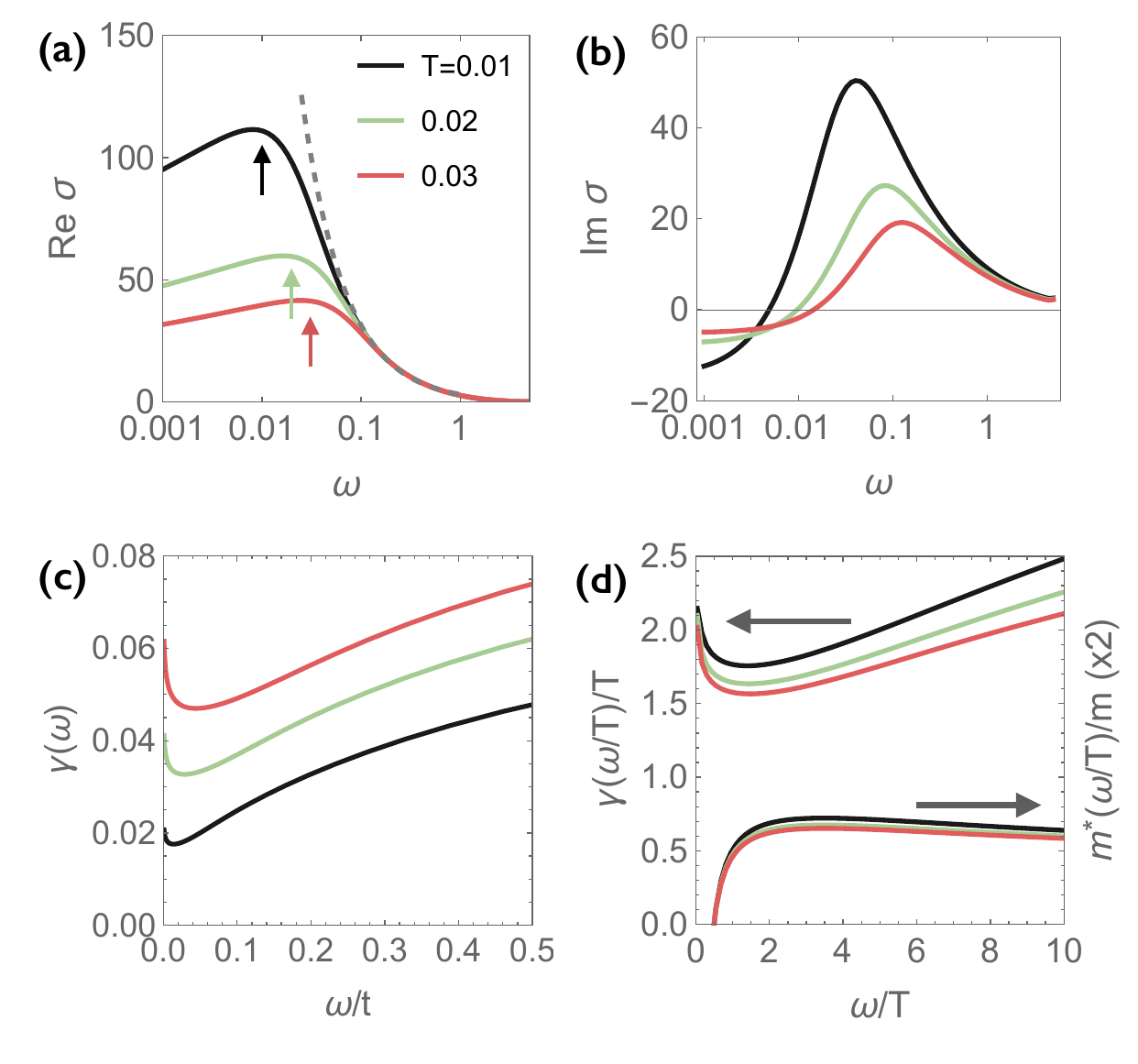}
    \caption{(a) Real part of the optical conductivity obtained from Eq. (\ref{eq:relation}) using the subdiffusive ansatz  Eq, (\ref{eq:sub}) with $\alpha=0.9$. The arrows indicate the Planckian scale $k_BT/\hbar$. The dashed line is $1/\omega^\alpha$. (b) Imaginary part  of the optical conductivity. (c) and (d) illustrate the corresponding extended Drude analysis, analogous to Fig. 3(a,b) of the main text.   }
    \label{fig:sub}
\end{figure}

In practice we take 
\begin{equation}
\label{eq:sub}
    \Delta X^2(t)/N =  \frac{v^2t^2}{1+(t/2\tau)^{2-\alpha}}
\end{equation}    
with $\alpha<1$ and insert it in Eq. (\ref{eq:relation}) in order to obtain the optical conductivity. 
For $\alpha=1$, Eq. (\ref{eq:sub}) describes ballistic then diffusive motion, providing an accurate analytical representation of the function $\Delta X^2(\mathrm{t})_{Drude}$ obtained through Eq. (\ref{eq:relinvclas}) (one can verify that applying Eq. (\ref{eq:relation}) in this case closely matches the Drude shape). If instead subdiffusive motion is considered, $\alpha<1$, one obtains the optical conductivity illustrated in Fig. \ref{fig:sub}(a,b). Importantly, the real part of the optical conductivity now develops a  displaced Drude peak, located at a frequency $\omega_{\mathrm{peak}}\approx k_BT/\hbar$, and the imaginary part becomes negative at low frequencies (note the similarity with Fig1(a,b) of Ref. [5]).
The high frequency behavior of the optical conductivity becomes  $Re \ \sigma \sim 1/\omega^\alpha$ (dashed line). 

Fig. \ref{fig:sub}(c,d) show the optical scattering rate and the effective mass renormalization obtained for subdiffusive motion, to be compared with Fig. 3(a,b) of the main text (here again, the results are very similar to those of Ref. [5], Fig.1(c,d)). 
 
\vskip 2cm 

[45] C. Berthod, J. Mravlje, X. Deng, R. \v{Z}itko, D. van der Marel, and A. Georges, Non-Drude universal scaling laws for the
optical response of local Fermi liquids, Phys. Rev. B 87, 115109 (2013)


\end{document}